\documentclass[aps,pra,twocolumn,showpacs,superscriptaddress,nofootinbib]{revtex4-1}  % for review and submission

\usepackage{graphicx}  % needed for figures
\usepackage{dcolumn}   % needed for some tables
\usepackage{amssymb}   % for math
\usepackage[caption=false]{subfig}
\usepackage{amsmath}
\usepackage{siunitx}

\DeclareMathOperator{\Tr}{Tr}

\makeatother

\begin{document}

\title{Experimental and theoretical investigation of a multi-mode cooling scheme using multiple EIT resonances}
\author{Nils Scharnhorst}
	\affiliation{QUEST Institute for Experimental Quantum Metrology, Physikalisch-Technische Bundesanstalt, 38116 Braunschweig,~Germany}
	\affiliation{Institut f{\"u}r Quantenoptik, Leibniz Universit{\"a}t Hannover, 30167 Hannover,~Germany}
\author{Javier Cerrillo}     
	\affiliation{Faculty II Mathematics and Natural Sciences, Institute for Theoretical Physics, Technische Universit\"at, 10623 Berlin,~Germany}
\author{Johannes Kramer}
	\affiliation{QUEST Institute for Experimental Quantum Metrology, Physikalisch-Technische Bundesanstalt, 38116 Braunschweig,~Germany}
\author{Ian D. Leroux}
	\altaffiliation[Current adrdress: ]{National Research Council Canada, Ottawa, Ontario, K1A 0R6, Canada}
	\affiliation{QUEST Institute for Experimental Quantum Metrology, Physikalisch-Technische Bundesanstalt, 38116 Braunschweig,~Germany}
\author{Jannes B. W{\"u}bbena}
	\altaffiliation[Current adrdress: ]{Geo++ GmbH, 30827 Garbsen}
	\affiliation{QUEST Institute for Experimental Quantum Metrology, Physikalisch-Technische Bundesanstalt, 38116 Braunschweig,~Germany}
\author{Alex Retzker}
	\affiliation{Racah Institute of Physics, Hebrew University of Jerusalem, 91904 
	Jerusalem, Israel}
\author{Piet O. Schmidt}
	\affiliation{QUEST Institute for Experimental Quantum Metrology, Physikalisch-Technische Bundesanstalt, 38116 Braunschweig,~Germany}
	\affiliation{Institut f{\"u}r Quantenoptik, Leibniz Universit{\"a}t Hannover, 30167 Hannover,~Germany}
\date{\today}

%% ABSTRACT %%%%%%%%%%%%%%%%%%%%%%%%%%%%%%%%%%%%%%%%%%%%%%%%%%%%%%%%%%%%
\begin{abstract}
We introduce and demonstrate double-bright electromagnetically induced transparency (D-EIT) cooling as a novel approach to EIT cooling. By involving an additional ground state, two bright states can be shifted individually into resonance for cooling of motional modes of frequencies that may be separated by more than the width of a single EIT cooling resonance. This allows three-dimensional ground state cooling of a $^{40}$Ca$^+$ ion trapped in a linear Paul trap with a single cooling pulse. Measured cooling rates and steady-state mean motional quantum numbers for this D-EIT cooling are compared with those of standard EIT cooling as well as concatenated standard EIT cooling pulses for multi-mode cooling. Experimental results are compared to full density matrix calculations. We observe a failure of the theoretical description within the Lamb-Dicke regime that can be overcome by a time-dependent rate theory. Limitations of the different cooling techniques and possible extensions to multi-ion crystals are discussed.
\end{abstract}

\maketitle

%% MAIN TEXT %%%%%%%%%%%%%%%%%%%%%%%%%%%%%%%%%%%%%%%%%%%%%%%%%%%%%%%%%%%

%% Motivation %%%%%%%%
\section{Introduction}
\label{introduction}
Ground state cooling (GSC) of trapped ions and neutral atoms is a necessity in many quantum optics experiments such as quantum simulations \cite{porras_effective_2004, blatt_quantum_2012},  quantum state engineering \cite{monz_14-qubit_2011, leibfried_creation_2005, haffner_scalable_2005, johnson_ultrafast_2016, lo_spin-motion_2015}, quantum logic spectroscopy \cite{wan_precision_2014, wolf_non-destructive_2016, rosenband_frrequency_2008, Schmidt2005, hempel_entanglement-enhanced_2013, chou_preparation_2017}, and quantum logic clocks \cite{rosenband_frrequency_2008, Ludlow_ClocksReview} in order to minimize motion-induced errors and to obtain full quantum control over the system. Cooling performance is characterized by the cooling rate, the minimal achievable kinetic energy, and how many motional modes are cooled simultaneously. Fast cooling and thus short cooling times are important in all applications to reduce the associated dead time. This is particularly important in optical clocks where dead time directly leads to reduced stability through the Dick effect \cite{Ludlow_ClocksReview, poli_optical_2013, dick_local_1987}, limiting applications e.g. in relativistic geodesy \cite{vermeer_chronometric_1983, bjerhammar_relativistic_1985, lisdat_clock_2016}, or tests of fundamental physics \cite{rosenband_frrequency_2008, godun_frequency_2014, huntemann_improved_2014}. Standard techniques for GSC include sideband cooling \cite{diedrich_laser_1989, monroe_resolved-sideband_1995, roos_quantum_1999, vuletic_degenerate_1998, han_3d_2000, hamann_resolved-sideband_1998} and more recently, fast cooling via electromagnetically induced transparency (EIT) \cite{Morigi2000, roos_experimental_2000, lin_sympathetic_2013, kampschulte_eit-control_2012, Lechner2016_EIT}. 

EIT cooling \cite{Morigi2000, eschner_laser_2003} has become a standard GSC technique for trapped ions and atoms. It has been demonstrated for cooling of a single ion \cite{roos_experimental_2000}, for cavity-assisted EIT cooling of a neutral atom in a dipole trap \cite{kampschulte_eit-control_2012}, for sympathetic cooling of ion chains \cite{ lin_sympathetic_2013, Lechner2016_EIT}, and a quantum-gas microscope \cite{haller_single_2015}. Due to the narrow cooling resonance, standard EIT cooling restricts GSC to a narrow range of nearby vibrational modes in one single cooling pulse \cite{lin_sympathetic_2013, Lechner2016_EIT}. The larger the spectral-mode spread, the less efficient the cooling, potentially leading to heating of some modes. However, many quantum optics experiments such as quantum simulations of Ising models \cite{porras_effective_2004}, ion-strings \cite{ jurcevic_quasiparticle_2014, richerme_propagation_2014, Kiethe_Probing_2017}, and quantum logic spectroscopy \cite{ wan_precision_2014, wolf_non-destructive_2016, rosenband_frrequency_2008, Schmidt2005, hempel_entanglement-enhanced_2013, chou_preparation_2017} would benefit from a multi-mode GSC technique. EIT cooling is a natural extension of dark state cooling of atoms in free space \cite{aspect_laser_1988}. The analogy for trapped ions was first analyzed theoretically in \cite{Morigi2000, Morigi_Cooling_2003} and the main advantages over regular sideband cooling (SBC) were put forward in \cite{Morigi2000} and verified experimentally in \cite{roos_experimental_2000, schmidt-kaler_laser_2001}. Evers and Keitel \cite{evers_double-eit_2004} proposed a mechanism for the additional cancellation of the blue sideband through coupling to an extra level. Alternatively, the combination of EIT cooling with ground state driving was analyzed in \cite{cerrillo_fast_2010, Cerrillo2011}, where it was shown that a first order Lamb-Dicke coupling of the dark state to two, rather than one state also enables the cancellation of blue sideband scattering.

Here, we demonstrate a novel approach to EIT cooling, referred to as double-bright EIT (D-EIT) cooling, which allows simultaneous GSC of mode frequencies that may be separated by more than the width of a single EIT cooling resonance. 

The D-EIT scheme includes a third ground state that is coupled via an additional laser to an extra excited state. In real atomic systems heating processes due to off-resonant scattering from additional, unused, atomic levels typically dominate, so that the suppression of higher-order heating is of little benefit. Instead, we use the second freely tunable EIT resonance in the D-EIT level scheme to enable simultaneous cooling of spectrally-separated modes. At the same time, the scheme protects the dark state from decoherence due to off-resonant scattering and removes limitations due to other hot motional modes (spectator modes). We analytically characterize the performance of D-EIT cooling within second-order Lamb-Dicke perturbation theory in terms of final temperatures and cooling rates. Beyond the Lamb-Dicke limit, this characterization becomes unsuitable: on the one hand, the vibrational degrees of freedom are not strictly thermal and, on the other hand, the decay process does not follow a single exponential. Instead, the final Fock state occupation is used and a generalized cooling (time-dependent) rate is defined with help of a generalized fluctuation-dissipation theorem. This time-dependent rate evidences that the initial cooling stages are slower, and we ascribe this phenomenon to a depletion of the dark state induced by decoherence effects associated to the motional excitation and the higher coupling strength.

We experimentally demonstrate multi-mode ground-state cooling of all degrees of freedom of a single $^{40}$Ca$^+$ ion via D-EIT cooling within a single cooling pulse. The measured results on cooling rate and final temperature are compared to standard EIT cooling and sequential standard EIT cooling pulses for multi-mode cooling. All experimental results are complemented by full density matrix calculations.

The article is organized as follows: the experimental setup is introduced in Sec. \ref{experimental_setup} including the experimental sequence and the measurement method for the mean motional quantum number and cooling rate. The cooling schemes of single EIT and D-EIT are presented in Sec.~\ref{cooling_schemes}. Sec.~\ref{theory} discusses the theoretical description of the implemented cooling schemes, especially giving a description of cooling beyond the Lamb-Dicke perturbative limit. The experimental results are presented and discussed in Sec.~\ref{results}. Sec.~\ref{conclusion} summarizes and concludes the findings.

%% Setup %%%%%%%%%%%%%
\section{Experimental setup}\label{experimental_setup}

\begin{figure}
	\includegraphics[width=.99\linewidth]{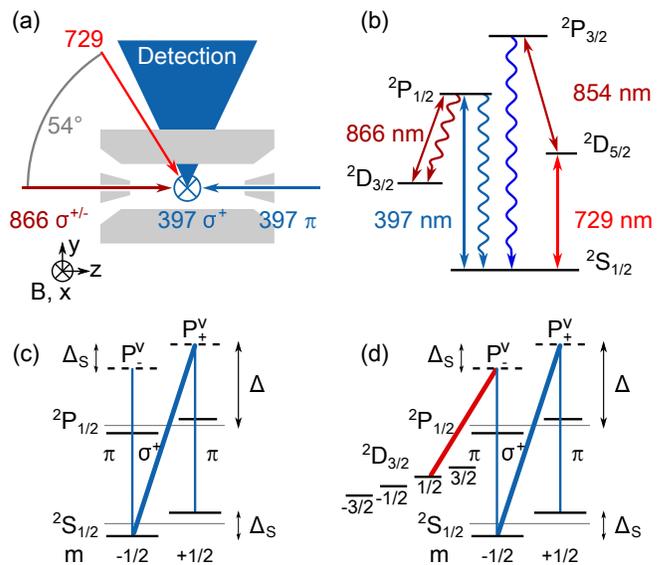}
	\caption{\label{levelslayout}(color online) (a) Schematic of the beam setup with respect to the trap and magnetic field $B$. (b) Energy levels of $^{40}$Ca$^+$. (c) Relevant levels for single EIT cooling. (d) Relevant levels for D-EIT. In (c), (d) the pump beam(s) are drawn with thicker lines.}
\end{figure}

\begin{figure}
	\includegraphics[width=1.\linewidth]{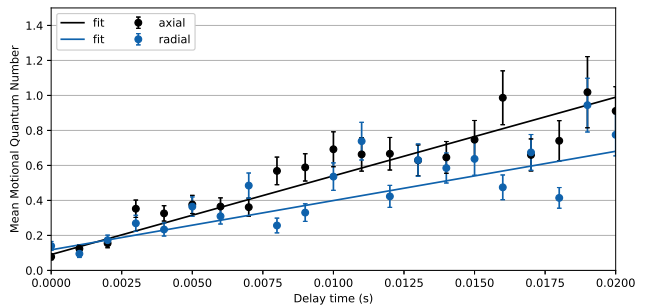}		
	\caption{(color online) Evolution of $\bar{n}(t)$ in radial and axial direction, measured after a delay time $t$ after D-EIT cooling, during which all lasers have been switched off. The background heating rate is derived from a linear fit of $\bar{n}(t)$ as \SI[parse-numbers=false]{(45.0\pm2.8)}{\per\second} in the axial direction and \SI[parse-numbers=false]{(28.2\pm3.4)}{\per\second} in the radial direction.\label{heatingrate}}
\end{figure}

A single $^{40}$Ca$^+$ ion is produced via ablation loading using a frequency-doubled pulsed Nd:YAG laser and ionized in a two-photon process \cite{sheridan_all-optical_2011, wubbena_photoionisation_2011, gulde_experimental_2003}. The $^{40}$Ca$^+$ ion is confined in a linear Innsbruck-style Paul trap as described in \cite{Dolezal2015_BBR, Wuebbena2014} with \SI{2.5}{mm} endcap-to-ion distance and \SI{0.8}{mm} radio-frequency (RF) electrode-to-ion distance, and hollow endcap electrodes providing axial laser access. The radially confining RF quadrupole field is generated by applying \SI{28.209}{MHz} to both pairs of diagonally opposite RF electrodes while the relative RF phase between those two pairs is near $\pi$. This symmetric driving of the RF electrodes minimizes axial micromotion, since in this configuration the ground RF potential coincides with the symmetry axis of the trap and the ground of the 3D DC quadrupole field generated by the endcap electrodes. With a loaded Q-factor of 270 and \SI{5}{W} of power fed to the helical resonator, the radial vibrational mode frequencies are measured to be \SI{2.552}{MHz} and \SI{2.540}{MHz} for a single $^{40}$Ca$^+$ ion, which implies a voltage between the RF trap electrodes of around \SI{600}{V}. The symmetry in the drive of the quadrupole field is deliberately slightly broken in order to lift the degeneracy of the radial modes. This represents a compromise between efficient radial cooling, which benefits from non-degenerate modes, and axial micromotion compensation, which benefits from symmetry. Radial micromotion is compensated via two pairs of DC electrodes running parallel to the RF electrodes. Axial confinement is provided by the two DC endcap electrodes. By applying a differential voltage between them, the ion can be placed at the point of minimal axial micromotion along the symmetry axis of the trap. The axial vibrational mode frequency $\nu_a$ is \SI{904.6}{kHz} with DC voltages of \SI{568.5}{\volt} and \SI{460}{\volt} applied to the endcap electrodes. For a single $^{40}$Ca$^+$ ion, micromotion is compensated to an overall residual second order fractional Doppler shift of less than $3\cdot10^{-18}$ via the sideband technique \cite{Berkeland1998-MM, keller_precise_2015}, corresponding to a modulation index $<0.05$. As shown in Fig.~\ref{heatingrate}, the background heating rate of the system is \SI[parse-numbers=false]{(45.0\pm2.8)}{\per\second} in the axial and  \SI[parse-numbers=false]{(28.2\pm3.4)}{\per\second} in the radial direction. These values are negligible compared to the cooling rates discussed in this work. Three pairs of field coils create a constant magnetic field oriented perpendicular to the axial direction with a field strength of \SI{416}{\micro T} (see Fig.~\ref{levelslayout}). An additional three pairs of field coils are used for active magnetic field stabilization. The overall field is measured in three orthogonal directions with a flux gate magnetic field probe \cite{magnetic_field_probe} and stabilized by feeding back on the current drivers of the coil. Along the quantization axis ($z$ direction), we achieve a suppression of \SI{10}{dB} for noise frequencies up to \SI{50}{Hz} and a residual rms field of \SI{3.3}{nT} in the frequency band from $30\:$Hz to $2\:$kHz.  

\begin{figure}
	\includegraphics[width=.99\linewidth]{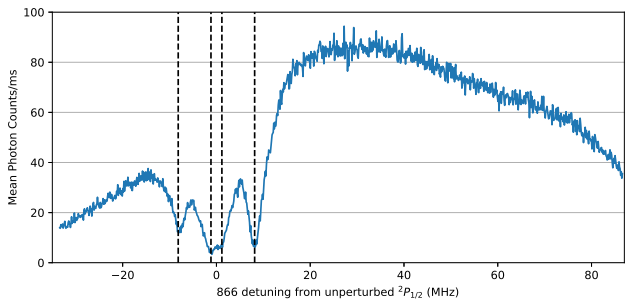}
	\caption{\label{Darkresonance}(color online) Fluorescence plot of dark resonances between \SI{866}{nm} laser and \SI{397}{nm} $\pi$ beam, where the latter is red-detuned by \SI{21.34}{MHz} from the unperturbed $^2P_{1/2}$ resonance. Dashed lines indicate theoretically expected frequencies of dark resonances for the given polarizations and magnetic field.}
\end{figure}

Fig.~\ref{levelslayout}(b) shows the relevant energy levels of $^{40}$Ca$^+$, connected by the available lasers \cite{toptica_lasers} labeled with their respective wavelength. As shown in Fig.~\ref{levelslayout}(a), two \SI{397}{nm} beams are available. One of them is purely radial and $\sigma^+$ polarized, the other one is directed along the axial direction and is $\pi$ polarized. The \SI{866}{nm} laser is also purely axial and $\sigma^{\pm}$ polarized. The \SI{729}{nm} beam has a projection onto the axial and radial directions. The \SI{729}{nm}, \SI{866}{nm}, and \SI{397}{nm} lasers are phase-locked to the same ultra-stable, cavity-stabilized reference using a high-bandwidth transfer-lock technique described in \cite{Scharnhorst2015_lock}. This ensures phase stability among the lasers, demonstrated by dark resonances between the \SI{866}{nm} and \SI{397}{nm} $\pi$ laser shown in Fig.~\ref{Darkresonance}. 

\subsection{Experimental sequence}
Every beam is switched and tuned in frequency individually via acousto-optical modulators (AOM). The timing and frequency of the optical pulses generated by the AOMs are controlled by an FPGA-based control-system \cite{schindler_PB_2008, Wuebbena2014}. All measurements use the same experimental pulse order:  For each data point the ion is first Doppler cooled for \SI{1}{\milli\second} via both \SI{397}{nm} lasers connecting the $^2S_{1/2}$ to $^2P_{1/2}$ levels while simultaneously applying the \SI{866}{nm} beam to act as a repumper. This prepares the ion close to the Doppler cooling temperature limit, corresponding to $\bar{n}\sim11.1$ in the axial and $\bar{n}\sim3.6$ in the radial modes assuming the natural linewidth of $\Gamma=2\pi\times$\SI{20.7}{MHz} of the $^2$S$_{1/2}$ to $^2$P$_{1/2}$ transition. After this first cooling step, EIT cooling is applied. This is followed by a low-power \SI{2}{\micro\second} optical pumping pulse of \SI{397}{nm} $\sigma^+$ and \SI{866}{nm} light preparing the population into the $|^2S_{1/2}, m_\mathrm{F}=1/2\rangle=|S_+\rangle$ state. The polarization purity of the \SI{397}{nm} $\sigma^+$ is better than $100:1$, which implies a measurement offset due to imperfect pumping and heating during the pump pulse of $\Delta\bar{n}\ll0.037$. This offset is negligible for our purposes. 
Information about the motional state of the ion is probed by mapping it onto the internal electronic states. For this purpose, we apply laser radiation at \SI{729}{nm} detuned to the red (blue) of the optical carrier by the oscillation frequency of the ion in the trap, driving red sidebands (RSB) (blue sidebands (BSB)) from $|S_+\rangle$ to $|^2D_{5/2}, m_\mathrm{F}=5/2\rangle=\left|\uparrow\right\rangle$ for a Rabi time $t_\mathrm{R}$. These pulses not only change the electronic state, but also decrease (increase) the motional quantum state. The final electronic states are discriminated by probing the $^2$S$_{1/2}$ to $^2$P$_{1/2}$ transition with \SI{397}{nm} $\pi$ light for \SI{250}{\micro\second}. During this time the ion's fluorescence is detected through imaging optics and a photo-multiplier tube. For each setting the sequence is repeated 250 times and from the resulting excitation statistics the population in the $\left|\uparrow\right\rangle$ logic state is inferred via the threshold technique \cite{roos_quantum_1999, hemmerling_novel_2012}. Typically, around \SI{100}{\per\milli\second} scattering events are detected if the ion is in the electronic ground state, and around \SI{0.1}{\per\milli\second} if the ion is in the $\left|\uparrow\right\rangle$ state. Every event with a scattering rate less than \SI{6.5}{\per\milli\second} is attributed to the ion being in the $\left|\uparrow\right\rangle$ state (dark state). The excitation probability is estimated as a fraction of trials in which the dark state was observed.

%% Methodology %%%%%%%
\subsection{Measurement of $\bar{n}$ and cooling rates}\label{methodology}
\begin{figure}
	\includegraphics[width=.99\linewidth]{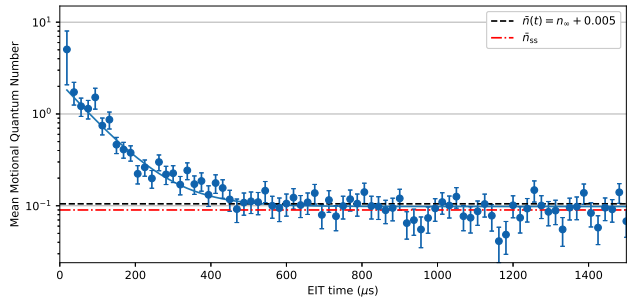}
	\caption{\label{tempsample}(color online) Typical dependency of $\bar{n}$ on cooling time, here for the axial direction after D-EIT cooling at $\Delta=3\Gamma$. The solid line shows the exponential fit. The horizontal dashed line indicates where the exponential fit approaches the fitted $\bar{n}$ to within $|n_\mathrm{\infty}-\bar{n}(t)|\leq0.005$, and the dash-dotted horizontal line represents the resulting $n_\mathrm{ss}$, which is calculated by averaging over all $\bar{n}(t)$ after the fit crosses below the dashed line.}
\end{figure}

For a single trapped particle a mean motional quantum number $\bar{n}$ can be assigned to each of its motional degrees of freedom (two radial modes, one axial mode). In case of a thermal distribution, the $\bar{n}$ corresponds to a temperature. The temporal evolution $\bar{n}(t)$ after applying GSC for time $t$ is obtained by sideband thermometry \cite{turchette_nbar_2000}. For this, the excitation of RSB and BSB at a given Rabi time are measured as a function of cooling time. The temperature is then computed as \[\bar{n}(t)=RSB(t)/(BSB(t)-RSB(t))\]from the sideband excitations. A typical temperature measurement with corresponding fit is shown in Fig.~\ref{tempsample}. From the temperature measurements the steady-state mean motional quantum number $n_\mathrm{ss}$ is obtained in the following way. First, the data points are fitted with an exponential function of the form $\bar{n}(t)=A\exp({-R t})+n_\mathrm{\infty}$, with $R$ being the cooling rate. Due to the non-exponential temporal evolution, the residuals between fit and the temperature measurements for $t\rightarrow\infty$ can end up not being normally distributed around $n_\mathrm{\infty}$. Therefore, a cut-off time $t_\mathrm{cut}$ is introduced after which the exponential fit approaches the fitted $\bar{n}$ to within $|n_\mathrm{\infty}-\bar{n}(t)|\leq0.005$. We then determine the steady-state mean motional quantum number $n_\mathrm{ss}$ by averaging over all data points with $t>t_\mathrm{cut}$. This procedure ensures small uncertainties in the determined $n_\mathrm{ss}$ despite the non-exponential evolution of the measured $\bar{n}(t)$. 

For temperatures $n(t)\gg\bar{n}_\mathrm{ss}$, the measurement SNR of $\bar{n}$ is improved by first fitting the sideband excitations individually with an exponential function and then deriving the $\bar{n}(t)$ from the fitted $RSB(t)$, $BSB(t)$. A further improvement of the SNR of  $\bar{n}$ is achieved by probing the sidebands at a time corresponding to the $\pi$-time of a Doppler-cooled Rabi flop. This shorter $\pi$-time and the additional fitting step for the sidebands has been chosen to determine the cooling rate $R$.

For comparison with the simulations, Rabi frequencies of the cooling beams (\SI{397}{nm} $\pi$ and $\sigma^+$) have been calibrated in the following way. A part of the power of each beam is picked off before the beams enter the vacuum chamber and is monitored with a photo diode (PD) for active intensity stabilization by feeding back on the AOM RF drive power using a digital sample \& hold stabilization circuit. The induced Stark shift $\Delta_{\mathrm{Stark}}$ on the state $|S_\pm\rangle$ is deduced by taking the difference between the carrier resonance frequency of the logic transition $|S_+\rangle\:(|S_-\rangle)\leftrightarrow\left|\uparrow\right\rangle$ in case of the \SI{397}{nm} $\pi$ ($\sigma^+$) beam with and without one of the cooling beams being on during the logic pulse. Doing so for different cooling laser powers, a linear dependence between Stark shift and cooling laser voltage on the PD is obtained. The Rabi frequency can then be calculated using the relationship $\Omega_\mathrm{R}=\sqrt{4|\Delta|\Delta_\mathrm{Stark}}$. The \SI{866}{nm} Rabi frequency is a free parameter in the simulation which is chosen to fit the experimental results.

\begin{figure}
	\includegraphics[width=.5\linewidth]{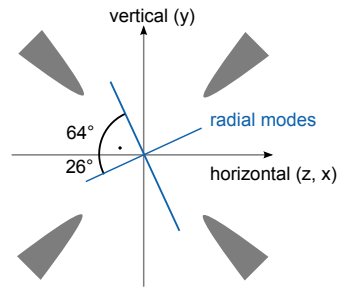}
	\caption{\label{modorientation}(color online) Cross section through the center of the ion trap, showing the orientation of radial modes with respect to RF electrodes, schematically represented as gray shaded areas.}
\end{figure}

As shown in Fig.~\ref{levelslayout}(a) all cooling beams run in the same plane to which the \SI{729}{nm} logic laser has an angle $\alpha=(54.7\pm5.1)^{\circ}$. The two radial modes are orthogonal with respect to each other. They are rotated around the trap axis such that the modes have a $(26\pm10)^{\circ}$ and $(-64\pm10)^{\circ}$ angle with respect to the plane of the cooling lasers, see Fig.~\ref{modorientation}. The rotation angle has been deduced from the different sideband Rabi-frequencies for the two radial modes for the same logic laser intensity. In the radial direction the temperature is obtained from the mode that has a lower projection onto the plane spanned by the cooling beams. The other radial mode has a higher projection onto this plane, resulting in more efficient cooling of this mode. The \SI{12}{kHz} spectral spacing between the two radial modes is small compared to all cooling resonance linewidths.

\section{Cooling schemes}\label{cooling_schemes}

\begin{figure}
	\includegraphics[width=.99\linewidth]{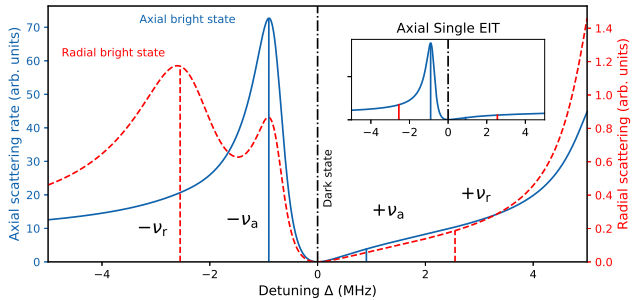}
	\caption{\label{EITscat}(color online) Simulated scattering rate during D-EIT cooling versus detuning $\Delta$ from the virtual level $P_+$. Vertical straight lines indicate motional frequencies. The inset shows the single EIT spectrum in the respective direction optimized for axial cooling. D-EIT has a bright state at both radial and axial frequency, while for single EIT there is only one bright state. The position of the radial (axial) bright state is dominated by the Rabi frequency of the \SI{397}{nm} $\sigma$ (\SI{866}{nm} $\sigma^-$) beam.}
\end{figure}

\subsection{Single EIT cooling}\label{single_EIT}
EIT cooling is a technique based on a $\Lambda$-like level scheme with two ground states and one shared excited state \cite{Morigi2000}. Fig.~\ref{levelslayout}(c) shows the relevant levels and beams for single EIT cooling in $^{40}$Ca$^+$. Here, the ground states are the $|^2S_{1/2}, m_\mathrm{F}=\pm1/2\rangle=|S_\pm\rangle$ levels and the excited state is chosen to be the $|^2P_{1/2}, m_\mathrm{F}=1/2\rangle=|P_+\rangle$ state. If lasers connecting the respective ground states are brought into two-photon resonance near the excited state, a dark resonance is generated, as shown for a red-detuned resonance in Fig.~\ref{Darkresonance} and for a blue detuned resonance in the simulated scattering spectrum in the inset of Fig.~\ref{EITscat}. Each dark resonance is associated with a bright state that is theoretically described by a Fano-profile \cite{Janik_Fano}. Its spectral position (shift) with respect to the dark resonance depends on the Rabi frequency of the two lasers forming the dark resonance. Typically, one of the two lasers is much stronger (pump) than the other one (probe) such that the shift of the bright state is dominated by the pump beam. In our setup the \SI{397}{nm} $\sigma^+$ beam is used as pump and the \SI{397}{nm} $\pi$ serves as probe beam. Pump and probe are brought into resonance on a virtual energy level $P_\mathrm{+}^v$ which is blue detuned by $\Delta$ with respect to the unperturbed $^2P_{1/2}$ level. That choice of probe beam reduces the effect of off-resonant scattering on the $|S_-\rangle$ to $|P_-\rangle$ transition \cite{roos_experimental_2000}. Our beam geometry allows cooling of the axial or radial modes by shifting the bright state to the respective mode frequency. Due to the large mode spacing, a single EIT pulse can cool either the axial or the two radial modes efficiently, but not all three at once. All modes can be cooled using single EIT by applying two sequential pulses, which is hereafter referred to as  ``single EIT with precooling'' or ``3D EIT cooling''.

\subsection{Double-bright EIT cooling}\label{double_EIT}
In D-EIT cooling, the $\Lambda$ scheme of single EIT cooling is extended to a double $\Lambda$-shaped level scheme, as shown in Fig.~\ref{levelslayout}(d). This is accomplished by introducing a \SI{866}{nm} beam as an additional pump connecting the $|^2D_{3/2}, m_\mathrm{F}=1/2\rangle=|D_+\rangle$ to the virtual $P_-^v$ level. It contributes to the dark resonance suppression of carrier scattering and creates a second bright state together with the \SI{397}{nm} $\pi$ beam coupling $|S_-\rangle$ to $|^2P_{1/2}, m_\mathrm{F}=-1/2\rangle=|P_-\rangle$ as shown in Fig.~\ref{EITscat}, hence the name double-bright EIT. For our beam geometry, the second bright state is formed by purely axial beams and it is brought into resonance with the axial motional sideband by the Stark-shift via the \SI{866}{nm} pump beam. The bright state formed by the \SI{397}{nm} $\pi$ and $\sigma^+$ beams is shifted into resonance with the radial sidebands. The shift of the two bright states can be chosen independently by adjusting the Rabi frequency of the respective pump beam accordingly. This way, both radial and axial modes can be cooled simultaneously, despite their frequency splitting of more than \SI{1.5}{\mega\hertz}. Furthermore, the off-resonant coupling of the \SI{397}{nm} $\pi$ transition to the $|P_-\rangle$ level is made dark via coupling to the $^2$D$_{3/2}$-manifold, which reduces off-resonant scattering.

The plot of the scattering rate for D-EIT cooling (Fig.~\ref{EITscat}) exhibits higher rates on the side of positive detuning compared to that of single EIT cooling. This is a result of the onset of one of the additional scattering resonances created by couplings between other levels of the $^2D_{3/2}$- and $^2P_{1/2}$-manifolds. In the case of single EIT cooling those additional resonances are present as well, but further detuned from the dark state.

%% Theory %%%%%%%%%%%%
\section{Theory}\label{theory}
\subsection{Lamb-Dicke Theory}

In order to theoretically model the performance of both cooling schemes
and to provide predictions for $R$ and $n_\mathrm{ss}$, we firstly apply a
commonly used approach involving adiabatic elimination of the electronic
degrees of freedom in the Lamb-Dicke limit \cite{Cirac1992}. In this
limit, two competing processes dictate the cooling rate and the steady-state
temperature, which are the diffusion due to spontaneous emission and
the coordinated effect of sideband excitation and dissipation. The
first one is known as \emph{carrier scattering} and contributes with
a diffusion term of the form
\[
D=\sum_{e}p_{e}\sum_{g}\tilde{\eta}_{eg}^{2}\gamma_{eg},
\]
where the sum is over all excited states $e$ and all ground states
$g$, $p_{e}$ denotes the steady state population of the excited
state, $\tilde{\eta}_{eg}$ is the effective Lamb-Dicke parameter
associated with the $e$ to $g$ transition and $\gamma_{eg}$ its
scattering rate. The second one is quantified in terms of the steady-state,
two-time correlation function in frequency domain
\[
S\left(\omega\right)=\int_{0}^{\infty}e^{-i\omega t}\left\langle \hat{\sigma}_{\eta}\left(t\right)\hat{\sigma}_{\eta}\right\rangle _{\mathrm{ss}}dt,
\]
of the operator $\hat{\sigma}_{\eta}$ corresponding to the electronic
part of the first order term of the Lamb-Dicke expansion of the Hamiltonian.
Here we make use of the averaging $\left\langle \bullet\right\rangle _{\mathrm{ss}}=\Tr\left\{ \bullet\rho_{\mathrm{ss}}\right\} $
with respect to the zeroth-order steady-state electronic density matrix
$\rho_{\mathrm{ss}}$. Its evaluation at the oscillator frequency
$\nu$ provides the heating ($A_{+}$) and cooling ($A_{-}$) rates
\[
A_{\pm}\equiv2\Re\left[S\left(\pm\nu\right)+D\right],
\]
where $\Re$ stands for the real part. These enter the rate equation
$\frac{d}{dt}\vec{p}=\mathbf{A}\vec{p}$ governing the population
dynamics $\vec{p}^{T}=\left[p_{0}\left(t\right),\dots,p_{n}(t),\dots\right]$
in the Fock space of the oscillator
\begin{alignat}{1}
\frac{d}{dt}p_{n}= & -\left[\left(n+1\right)A_{+}+nA_{-}\right]p_{n}\label{eq:RateLD}\\
& +\left(n+1\right)A_{-}p_{n+1}+nA_{+}p_{n-1},\nonumber 
\end{alignat}
corresponding to a matrix of the form
\begin{alignat}{1}
\mathbf{A}_{j,k}= & -\left[\left(j+1\right)A_{+}+jA_{-}\right]\delta_{j,k}\nonumber \\
& +kA_{-}\delta_{j+1,k}+jA_{+}\delta_{j,k+1}.\label{eq:Amatrix}
\end{alignat}
The predicted steady state probability distribution
\begin{equation}
p_{n}^{\mathrm{ss}}=\left(\frac{A_{+}}{A_{-}}\right)^{n}p_{0}^{\mathrm{ss}},\label{eq:pthermal}
\end{equation}
corresponds to a thermal state of temperature $T_{\mathrm{ss}}=\frac{\nu}{k_{\mathrm{B}}\ln\frac{A_{-}}{A_{+}}}$
, average occupation $n_{\mathrm{ss}}=\frac{A_{+}}{A_{-}-A_{+}}$
and ground state occupation $p_{0}^{\mathrm{ss}}=\frac{A_{-}-A_{+}}{A_{-}}$.
Additionally, the approach to the steady state is governed by the
cooling rate $R=A_{-}-A_{+}$.

\subsubsection{Single EIT Cooling}

The traditional concept for EIT cooling arising in the context of
a Raman laser configuration \cite{Morigi2000} is such that, within
the regime of validity of the Lamb-Dicke perturbative expansion, the
presence of a dark state $\left|-\right\rangle $ eliminates the carrier
scattering $D$ and additionally simplifies the mathematical derivation
of the absorption spectrum $S\left(\omega\right)$. The dark state
arising in our configuration is $\left|-\right\rangle =\frac{1}{\Omega}\left(\Omega_{\pi}\left|S_{-}\right\rangle -\Omega_{\sigma}\left|S_{+}\right\rangle \right)$.
It involves a superposition of the two ground states with $\Omega\equiv\sqrt{\Omega_{\pi}^{2}+\Omega_{\sigma}^{2}}$,
and $\Omega_{\pi,\sigma}$ the Rabi frequencies of the 397 nm $\pi$
and $\sigma$ polarized lasers respectively, whereas the cooling operator
may be defined as $\hat{\sigma}_{\eta}=\left(\left|\eta\right\rangle \left\langle P_{+}\right|+H.c.\right)$,
where $\left|\eta\right\rangle =\frac{1}{2}\left(\eta_{\sigma}\Omega_{\sigma}\left|S_{-}\right\rangle +\eta_{\pi}\Omega_{\pi}\left|S_{+}\right\rangle \right)$
is an unnormalized state involving the Lamb-Dicke parameters $\eta_{\pi,\sigma}$
of the two lasers along the relevant axis as shown in Fig.~\ref{levelslayout}(a).

In this case, the real part of the absorption spectrum as derived by means of the quantum
regression theorem \cite{Gardiner2004} becomes \cite{Morigi2000, Morigi_Cooling_2003}
\begin{alignat}{1}
\Re \left[S\left(\omega\right) \right] & =\frac{\eta^{2}}{4}\left(\frac{\Omega_{\pi}\Omega_{\sigma}}{\Omega}\right)^{2}\frac{\gamma}{\left(\omega-\Delta-\cfrac{\Omega^{2}}{4\omega}\right)^{2}+\gamma^{2}},\label{eq:SEIT}
\end{alignat}
where $\eta=\eta_{\pi}-\eta_{\sigma}$ and $\gamma$ is the decay
rate of the $\left|P_{+}\right\rangle $ state. Eq.(\ref{eq:SEIT})
vanishes for $\omega=0$ and contains two maxima at $\omega_{\pm}=\frac{\Delta}{2}\pm\frac{1}{2}\sqrt{\Delta^{2}+\Omega^{2}}$,
corresponding to the frequencies of the dressed excited state $\left|P_{+}\right\rangle $
and bright state $\left|+\right\rangle =\frac{1}{\Omega}\left(\Omega_{\sigma}\left|S_{-}\right\rangle +\Omega_{\pi}\left|S_{+}\right\rangle \right)$
respectively. By means of a large blue detuning, the maximum of the
dressed bright state can be adjusted to the relevant oscillator frequencies
$\nu_{a,r}=-\omega_{-}\simeq\frac{\Omega^{2}}{4\Delta}$ so that $A_{+}\ll A_{-}$
and an optimal cooling rate \cite{Morigi2000, Morigi_Cooling_2003}
\begin{equation}
R^{EIT}\simeq\frac{\eta^{2}}{2\gamma}\left(\frac{\Omega_{\pi}\Omega_{\sigma}}{\Omega}\right)^{2},\label{eq:REIT}
\end{equation}
and a final occupation number 
\begin{equation}
n_{\mathrm{ss}}^{EIT}\simeq\frac{\gamma^{2}}{4\Delta^{2}+\gamma^{2}}+n_{\pi},
\end{equation}
may be achieved, where $n_{\pi}$ corresponds to the heating contribution
of the spurious 397 $\pi$ coupling. This contribution is approximately
of the form $n_{\pi}\simeq\frac{2\gamma^{2}}{4\Delta^{2}+\gamma^{2}}\left(\frac{\Omega_{\pi}}{\Omega}\right)^{2}$
in the limit $\Omega_{\pi}\ll\Omega_{\sigma}$ of highly asymmetric
laser intensities used in the implementation of EIT in $^{40}\mathrm{Ca}^{+}$.
The dependence on $\Omega_{\pi}$ clarifies the necessity of this
limit to minimize the spurious coupling between $\left|S_{-}\right\rangle $
and $\left|P_{-}\right\rangle $ that depletes the dark state $\left|-\right\rangle $.
Although the condition maximizes Eq.(\ref{eq:REIT}) for constant
$\Omega$, it implies $R^{EIT}\simeq\frac{\eta^{2}\Omega_{\pi}^{2}}{2\gamma}$,
so that the probe laser establishes the order of magnitude of the
rate. Since $\Omega\simeq\Omega_{\sigma}$, the pump laser controls
the position of the dressed state and, by the relations above, $R^{EIT}\ll\frac{2\Delta}{\gamma}\eta^{2}\nu_{a,r}$.
Therefore, this configuration fundamentally limits the rates that
can be achieved for a given detuning. The additional laser coupling
proposed in double bright EIT circumvents this limitation and allows
EIT cooling with higher laser intensities. However, for the experimental parameter range investigated here, the contribution of $n_\pi$ is typically on the order of $n_\pi\lesssim 0.02$ and therefore negligible.

\subsubsection{Double-bright EIT cooling}

In our system, judicious choice of the detunings allows for six possible
implementations of a double-$\Lambda$-Raman configurations. The choice
involving the $\left|D_{+}\right\rangle $ level generates the following
dark state 
\begin{alignat*}{1}
\left|\sim\right\rangle  & =\frac{1}{\Omega_{-}^{2}}\left(\Omega_{D}\Omega_{\sigma}\left|S_{+}\right\rangle -\Omega_{D}\Omega_{\pi}\left|S_{-}\right\rangle +\Omega_{\pi}^{2}\left|D_{+}\right\rangle \right),
\end{alignat*}
with $\Omega_{-}^{4}=\Omega_{D}^{2}\Omega_{\sigma}^{2}+\Omega_{D}^{2}\Omega_{\pi}^{2}+\Omega_{\pi}^{4}$
and $\Omega_{D}$ the Rabi frequency associated to the 866 nm laser.
The cooling operator has the form
\begin{alignat*}{1}
\hat{\sigma}_{\eta} & =\frac{1}{2}\Big[i\left(\eta_{\pi}\Omega_{\pi}\left|S_{-}\right\rangle +\eta_{D}\Omega_{D}\left|D_{+}\right\rangle \right)\left\langle P_{-}\right|+\\
& i\left(\eta_{\sigma}\Omega_{\sigma}\left|S_{-}\right\rangle +\eta_{\pi}\Omega_{\pi}\left|S_{+}\right\rangle \right)\left\langle P_{+}\right|+H.c.\Big],
\end{alignat*}
where only the ground states with non-vanishing overlap with $\left|\sim\right\rangle $
are considered. In particular, $\hat{\sigma}_{\eta}\left|\sim\right\rangle =-i\left|E\right\rangle $
with the unnormalized state
\[
\left|E\right\rangle =\frac{\Omega_{\pi}\Omega_{D}}{2\Omega_{-}^{2}}\Big[\Omega_{\pi}\left(\eta_{D}-\eta_{\pi}\right)\left|P_{-}\right\rangle +\Omega_{\sigma}\left(\eta_{\pi}-\eta_{\sigma}\right)\left|P_{+}\right\rangle \Big].
\]

As in the case of single EIT, the presence of the dark state facilitates
the analytical calculation of the scattering spectrum, which can be
expressed in the form of a matrix inversion (see Appendix~\ref{Supplementary_scattering})
\[
S\left(\omega\right)=i \left\langle E\right|\left(\begin{array}{cc}
-a\left(\omega\right) & \frac{\Omega_{\pi}\Omega_{\sigma}}{4\omega}\\
\frac{\Omega_{\pi}\Omega_{\sigma}}{4\omega} & -b\left(\omega\right)
\end{array}\right)^{-1}\left|E\right\rangle.
\]
It contains the denominator
of a cooling spectrum exhibiting two EIT features (analogous to double
EIT cooling as described in \cite{evers_double-eit_2004})
\begin{alignat*}{1}
a\left(\omega\right) & =\omega-\Delta-\frac{1}{3}\Delta_{\mathrm{s}}+i\gamma-\frac{\Omega_{\pi}^{2}+\Omega_{D}^{2}}{4\omega}\\
& -\frac{3\Omega_{D}^{2}}{4\omega+4\frac{4}{5}\Delta_{\mathrm{s}}},
\end{alignat*}
and, what is more interesting, the denominator of a cooling spectrum
with three EIT features
\begin{alignat*}{1}
b\left(\omega\right) & =\omega-\Delta-\frac{2}{3}\Delta_{\mathrm{s}}+i\gamma-\frac{\Omega_{\pi}^{2}+\Omega_{\sigma}^{2}}{4\omega}\\
& -\frac{\Omega_{D}^{2}}{4}\left(\frac{3}{\omega+\frac{3}{5}\Delta_{\mathrm{s}}}+\frac{1}{\omega+\frac{7}{5}\Delta_{\mathrm{s}}}\right).
\end{alignat*}
This triple EIT structure has never been described in the literature
and is a novel aspect of our cooling scheme. Given the determinant
of the matrix

\[
\mathcal{D}(\omega)=a\left(\omega\right)b\left(\omega\right)-\frac{\Omega_{\pi}^{2}\Omega_{\sigma}^{2}}{16\omega^{2}},
\]
the final form of the scattering spectrum becomes
\begin{alignat*}{1}
S\left(\omega\right)= & \Im\frac{\Omega_{\pi}^{2}\Omega_{D}^{2}}{4\Omega_{-}^{4}\mathcal{D}\left(\omega\right)}\Big[\Omega_{\pi}^{2}\left(\eta_{D}-\eta_{\pi}\right)^{2}b\left(\omega\right)\\
& +\Omega_{\sigma}^{2}\left(\eta_{\pi}-\eta_{\sigma}\right)^{2}a\left(\omega\right)\\
& -\Omega_{\sigma}^{2}\Omega_{\pi}^{2}\left(\eta_{\pi}-\eta_{\sigma}\right)\left(\eta_{D}-\eta_{\pi}\right)\frac{1}{2\omega}\Big],
\end{alignat*}
where $\Im$ stands for the imaginary part.

The values of the Lamb-Dicke parameters control whether triple or
double EIT cooling takes place, or a combination of both. The radial
degrees of freedom are easiest to analyze since, due to the beam geometry
{[}see Fig.~\ref{levelslayout}(a){]}, $\eta_{\pi}=\eta_{D}=0$ and
a modified triple EIT term survives
\[
\Re\left[S^{r}\left(\omega\right)\right]=\frac{\eta_{\sigma}^{2}}{4}\frac{\Omega_{\pi}^{2}\Omega_{D}^{2}\Omega_{\sigma}^{2}}{\Omega_{-}^{4}}\Im\left[\frac{1}{b\left(\omega\right)-\frac{\Omega_{\pi}^{2}\Omega_{\sigma}^{2}}{16\omega^{2}a\left(\omega\right)}}\right].
\]
 On the other hand, the axial degree of freedom is characterized by
$\eta_{\sigma}=0$, and
\[
\Re\left[S^{a}\left(\omega\right)\right]=\frac{\eta_{\pi}^{2}}{4}\frac{\Omega_{\pi}^{2}\Omega_{D}^{2}}{\Omega_{-}^{4}}\Im\left\{ \frac{\Omega_{\pi}^{2}\alpha\left[\alpha b\left(\omega\right)-\frac{\Omega_{\sigma}^{2}}{2\omega}\right]+\Omega_{\sigma}^{2}a\left(\omega\right)}{\mathcal{D}\left(\omega\right)}\right\} ,
 \]
with $\alpha=\frac{\eta_{D}-\eta_{\pi}}{\eta_{\pi}}\simeq1.46$.

The benefit of such a rich spectral structure is highlighted when
several vibrational modes need to be addressed. In this case, the
position of several maxima can be distributed along the spectrum in
such a way that all modes enjoy the condition $A_{+}\ll A_{-}$. As
with single EIT, this is achieved by adjusting the Stark shift of
the bright states to the appropriate frequency so that optimal rates
of up to
\begin{alignat*}{1}
R_{r,max}^{D-EIT} & \simeq\frac{\eta_{\sigma}^{2}}{2\gamma}\frac{\Omega_{\pi}^{2}\Omega_{D}^{2}\Omega_{\sigma}^{2}}{\Omega_{-}^{4}},\\
R_{a,max}^{D-EIT} & \simeq\frac{\eta_{\pi}^{2}}{2\gamma}\frac{\Omega_{\pi}^{2}\Omega_{D}^{2}}{\Omega_{-}^{4}}\max\left(\Omega_{\pi}^{2}\alpha^{2},\Omega_{\sigma}^{2}\right),
\end{alignat*}
may be accessed. Stark shifts are in general proportional to the square
of the relevant Rabi frequency and inversely proportional to the detuning
of the shifting beam. Since Stark shifts need to be kept fixed in
this cooling configuration, cooling rates are roughly proportional
to the detuning of the laser configuration $R_{r,a;max}^{D-EIT}\simeq\frac{2\Delta}{\text{\ensuremath{\gamma}}}\eta_{\sigma,\pi}^{2}\nu_{a,r}$.
For the final occupation
\[
n_{\mathrm{ss}}^{D-EIT}\simeq\frac{\gamma^{2}}{4\Delta^{2}+\gamma^{2}}\leq n_{\mathrm{ss}}^{EIT},
\]
may be achieved. In summary, D-EIT cooling allows cooling of two individually
adjustable mode ranges. Furthermore, as a consequence of the absence
of spurious carrier heating in D-EIT, smaller steady-state temperatures
may be achieved, while at the same time saturating the possible cooling
rate. These results strictly hold only within the Lamb-Dicke approximation,
and need to be modified to account for experimentally observed effects
beyond the Lamb-Dicke limit. Additionally, it is also possible to
adjust $\Delta_{s}$ to take advantage of the double and triple EIT
structures appearing in the scattering spectrum, such that the phenomenon
described in \cite{evers_double-eit_2004} can be exploited to reach vanishing occupation
within second-order perturbation theory.

\subsection{Beyond the Lamb-Dicke Limit}

The double $\Lambda$ configuration of D-EIT avoids the detrimental
coupling present in single EIT cooling, such that the Lamb-Dicke theory
predicts significantly higher rates with sufficient laser power. These
rates can become so large that the Lamb-Dicke approximation is no
longer fulfilled, and strong correlations between the motional and
the electronic degrees of freedom set in that effectively limit the
cooling capability. A realistic prediction of the behavior in this
limit requires a more advanced theory and, in practical terms, numerical
simulations of the full electronic and vibrational master equation.
In this section, we provide an interpretation of this limitation in
terms of linear response theory, which we first show to be equivalent
to the Lamb-Dicke limit, and then express deviations therefrom in
terms of generalized fluctuation-dissipation theorems. In general,
this treatment shows that the definition of cooling rate is ambiguous,
and a time dependent cooling rate is a more appropriate description.
This is a direct consequence of the build-up of non-negligible correlations
between the vibrational and electronic degrees of freedom that seem
to effectively reduce the decay rate of higher-lying Fock states.
Therefore, the first stages of cooling take longer, and the observed
timescales for cooling remain larger than the Lamb-Dicke prediction.
Additionally, the resulting vibrational steady state does not exhibit
a thermal distribution, so it is not appropriate to ascribe a temperature
to the measured $n_{\mathrm{ss}}$.

\subsubsection{Linear response theory}

In the Lamb-Dicke limit, laser cooling can be understood as a thermalization
process ensuing from contact of the oscillator with a thermal bath.
Linear response theory suffices to characterize this phenomenon.
From this perspective, it is possible to associate the rate at which
the cooling process takes place with the conductance of the thermal
environment. The fluctuation-dissipation theorem \cite{Kubo1966},
both in its quantum and classical versions, relates the thermal conductance
of the oscillator $\kappa$ to the fluctuations around equilibrium
\begin{equation}
\kappa=\frac{\partial J}{\partial T}=\frac{k_{\mathrm{B}}\beta^{2}}{2}\frac{d}{dt}\left\langle \left(\delta H\right)^{2}\right\rangle_{t=0} ,\label{eq:FDT}
\end{equation}
where $J=-\frac{d}{dt}\left\langle H\right\rangle $ is the thermal
current departing the oscillator and $\left\langle \left(\delta H\right)^{2}\right\rangle $
is the variance of energy fluctuations around equilibrium, with $H$
the Hamiltonian of the oscillator and $\beta=\frac{1}{k_{\mathrm{B}}T}$
the inverse temperature in units of the Boltzmann constant $k_{B}$.
This establishes a dynamical picture whereby small temperature deviations
$\Delta T$ from equilibrium generate a current well approximated
by Fourier's law $J=\kappa\Delta T$. The heat capacitance of the
oscillator $C=\frac{\partial\left\langle H\right\rangle }{\partial T}$
allows us to formulate a linear response equation
\begin{equation}
J=\frac{\kappa}{C}\Delta\left\langle H\right\rangle ,\label{eq:FourierLaw}
\end{equation}
which, by using $H=\hbar\nu\left(\hat{n}+\frac{1}{2}\right)$ for
an oscillator of frequency $\nu$ and number operator $\hat{n}$,
can be solved to find the dynamics of the occupation of the oscillator
$\bar{n}\equiv\left\langle \hat{n}\right\rangle $ in its approach
to the steady state 
\begin{equation}
\bar{n}\left(t\right)=\left(n_{0}-n_{\mathrm{ss}}\right)e^{-\frac{\kappa}{C}t}+n_{\mathrm{ss}},\label{eq:Nexp}
\end{equation}
where $n_{0}$ and $n_\mathrm{ss}$ correspond to the initial and final occupation
numbers respectively. 

In the following, we make use of the fluctuation-dissipation theorem to show $R=\frac{\kappa}{C}$, corresponding to the Lamb-Dicke cooling rate $R$.
The variance of the energy fluctuations at equilibrium $\left\langle \left(\delta H\right)^{2}\right\rangle $
can be computed with help of the two-time measurement probability
$p\left(m,n;t\right)$ of observing the oscillator at level $m$ when
it is at equilibrium and at level $n$ in a subsequent measurement
performed at a later time $t$
\[
\left\langle \left(\delta H\right)^{2}\right\rangle =\sum_{m=0}^{\infty}\sum_{n=-m}^{\infty}\left(\hbar\nu n\right)^{2}p\left(m,m+n;t\right).
\]
The probability $p\left(m,n;t\right)$ can be decomposed into the
probability $p_{m}^{\mathrm{ss}}$ for the oscillator to be in state
$m$ initially and the probability $\mathcal{E}_{n,m}\left(t\right)$
to perform a transition between states $m$ and $n$ in the time $t$.
The latter is also the dynamical map $\mathcal{E}\left(t\right)$
\cite{ref:rivas2011}, which maps the initial probability distribution
to the one at time $t$, $\vec{p}\left(t\right)=\mathcal{E}\left(t\right)\vec{p}\left(0\right)$.
Given Eq.(\ref{eq:RateLD}), the dynamical map is easily computed
as $\mathcal{E}\left(t\right)=\exp\left(\mathbf{A}t\right)$, with
$\mathbf{A}$ from Eq.(\ref{eq:Amatrix}). Therefore,

\begin{alignat}{1}
\frac{d}{dt}\left\langle \left(\delta H\right)^{2}\right\rangle_{t=0}  & =\lim_{t\rightarrow0}\frac{d}{dt}\sum_{m=0}^{\infty}\sum_{n=-m}^{\infty}\left(\hbar\nu n\right)^{2}\mathcal{E}_{n+m,m}\left(t\right)p_{m}^{\mathrm{ss}},\nonumber \\
& =\sum_{m=0}^{\infty}\sum_{n=-m}^{\infty}\left(\hbar\nu n\right)^{2}\mathbf{A}_{n+m,m}p_{m}^{\mathrm{ss}}.\label{eq:J2LR}
\end{alignat}
The form of the rate matrix (Eq.~\ref{eq:Amatrix}) restricts the sum over $n$ to values $\left|n\right|=1$, so that
\begin{alignat*}{1}
\frac{1}{\left(\hbar\nu\right)^{2}}\frac{d}{dt}\left\langle \left(\delta H\right)^{2}\right\rangle _{t=0}= & \sum_{m=0}^{\infty}\left[mA_{-}+\left(1+m\right)A_{+}\right]p_{m}^{\mathrm{ss}}\\
= & \left(A_{-}+A_{+}\right)n_{\mathrm{ss}}+A_{+}\\
= & 2\frac{A_{+}A_{-}}{A_{-}-A_{+}}.
\end{alignat*}
Therefore, given the heat capacity of the oscillator in its steady
state $C=k_{B}\left(\beta\hbar\nu\right)^{2}\frac{n_{\mathrm{ss}}}{p_{0}^{\mathrm{ss}}}$
the rate is confirmed to coincide with that of the linear response
prediction
\begin{equation}
\frac{\kappa}{C}=\frac{1}{\left(\hbar\nu\right)^{2}}\frac{d}{dt}\left\langle \left(\delta H\right)^{2}\right\rangle _{t=0}\frac{p_{0}^{\mathrm{ss}}}{2n_{\mathrm{ss}}}=A_{-}-A_{+}=R.\label{eq:RLR}
\end{equation}

\subsubsection{Non-linear response theory}

The linear response prediction only applies as long as the assumptions
of weak coupling and adiabatic separation of the timescales is satisfied.
This is expressed in our case by the condition $\eta_{i}\Omega_{i}\ll\nu$
for all laser couplings $i$, which sets an upper bound to the value
of the Rabi frequencies for which the Lamb-Dicke theory remains valid.
In practice, only the Lamb Dicke parameters and Rabi frequencies that
appear in the expression of the cooling rate $R$ play a role in the
strength of the dissipative dynamics and are therefore subject to
the upper bound. In consequence, for an EIT cooling rate of the form
$R\simeq\frac{\eta_{i}^{2}\Omega_{j}^{2}}{2\gamma}$, this implies
\begin{equation}
R\ll\frac{\nu^{2}}{2\gamma},\label{eq:LDlimit}
\end{equation}
which in our setting translates to rates much smaller than $10^{5}\mathrm{phonons}/\mathrm{s}$
for the axial degree of freedom and $10^{6}\mathrm{phonons}/\mathrm{s}$
for the radial degrees of freedom.

The behavior beyond these limits can only be predicted by considering
the full master equation for both the vibrational and electronic degrees
of freedom. In this situation, the reduced density matrix of the vibrational
degrees of freedom does not necessarily fit a thermal distribution
as in Eq.(\ref{eq:pthermal}) and the average occupation at the steady
state $n_\mathrm{ss}$, although well defined, is not associated to a single
temperature value. This is clear evidence of the breakdown of the
linear response limit, and the definition of a thermodynamically grounded
cooling rate as in Eq.(\ref{eq:RLR}) becomes troublesome.

In practical terms, the occupation number does not experience an exponential
decay as in Eq.(\ref{eq:Nexp}), but rather a decay whose rate depends
on the initial oscillator temperature and that varies as the steady
state is approached. This can be explored by introducing the concept
of an instantaneous rate $\frac{\partial J\left(t\right)}{\partial\left\langle H\right\rangle \left(t\right)}$.
For an initial thermal state of the oscillator, one may express the
derivative as the quotient \cite{inverse_map}.
\begin{equation}
\frac{\partial J\left(t\right)}{\partial\left\langle H\right\rangle \left(t\right)}=\frac{\kappa\left(t\right)}{C\left(t\right)},\label{eq:FourierExtended}
\end{equation}
where $C\left(t\right)=\frac{\partial\left\langle H\right\rangle \left(t\right)}{\partial T}$
is a generalized, time-dependent capacitance of the oscillator and
$\kappa\left(t\right)=\frac{\partial J\left(t\right)}{\partial T}$
is the corresponding generalized conductance. Whereas the fluctuation
dissipation theorem of the form Eq.(\ref{eq:FDT}) cannot be applied
in this case, it can be corrected by means of an additional term as
explained in \cite{Cerrillo2016} so that
\begin{alignat*}{1}
C\left(t\right) & =\frac{k_{B}\beta^{2}}{2}\left[\left\langle \left(\delta H\right)^{2}\right\rangle \left(t\right)-\left\langle H^{2}\right\rangle \left(t\right)\right],\\
\kappa\left(t\right) & =\frac{k_{B}\beta^{2}}{2}\frac{d}{dt}\left[\left\langle \left(\delta H\right)^{2}\right\rangle \left(t\right)-\left\langle H^{2}\right\rangle \left(t\right)\right].
\end{alignat*}
The first term on the right hand side is analogous to Eq.(\ref{eq:J2LR}),
so 
\begin{alignat*}{1}
\left\langle \left(\delta H\right)^{2}\right\rangle \left(t\right) & =\sum_{m=0}^{\infty}\sum_{n=-m}^{\infty}\left(\hbar\nu n\right)^{2}\mathcal{E}_{n+m,m}\left(t\right)p_{m}\left(T\right),
\end{alignat*}
with $p_{m}\left(T\right)$ the thermal distribution of temperature
$T$. The second term corrects the fluctuation dissipation theorem
and, as shown in \cite{Cerrillo2016}, corresponds to a computation
of the variance of the transferred energy
\begin{alignat*}{1}
\left\langle H^{2}\right\rangle \left(t\right) & =\sum_{m=0}^{\infty}\sum_{n=0}^{\infty}\left(\hbar\nu n\right)^{2}\mathcal{E}_{n,m}\left(t\right)p_{m}\left(T\right).
\end{alignat*}
Expression (\ref{eq:FourierExtended}) reduces to Eq.(\ref{eq:FourierLaw})
in the linear response limit since the fraction $\frac{\kappa\left(t\right)}{C\left(t\right)}$
becomes constant and independent of the initial temperature $T$.
Although the solution of Eq.(\ref{eq:FourierExtended}) does not have
a simple exponential decay form, the quotient $\frac{\kappa\left(t\right)}{C\left(t\right)}$
does provide us with some information about the timescale of the cooling
process.

The dynamical map $\mathcal{E}\left(t\right)$ can be well approximated
with help of the transfer tensors \cite{Cerrillo,Rosenbach2015},
which can be extracted from short timescale simulations of the global
master equations and contracted to provide values for the quotient
$R\left(t, \bar{n}_{0}\right)=\frac{\kappa\left(t\right)}{C\left(t\right)}$ to
very long times. 

\begin{figure}
	\includegraphics[width=1\columnwidth]{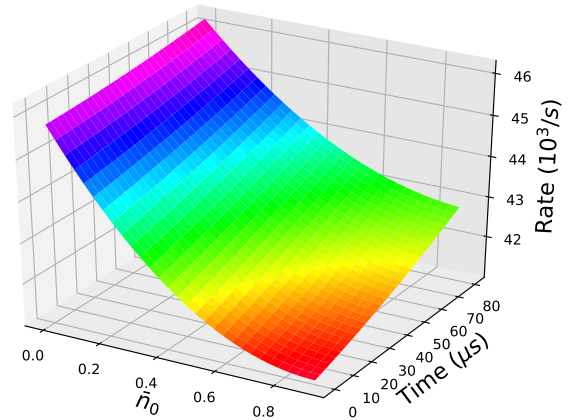}\protect\caption{\label{fig:Time-dependent-rate}Time dependent rate $R\left(t, \bar{n}_{0}\right)$
		in units of $10^{3}$ phonon$\cdot \mathrm{s}^{-1}$ for axial single EIT at
		$\Delta=3\Gamma$ as a function of the initial occupation number $\bar{n}_{0}$.
		The Lamb-Dicke prediction for the rate corresponds to $52.2\times10^{3}$
		phonon$\cdot \mathrm{s}^{-1}$, whereas the measured value is $38.2\times10^{3}$
		phonon$\cdot \mathrm{s}^{-1}$.}
\end{figure}

To theoretically evaluate the D-EIT cooling process outside the Lamb-Dicke regime, we performed numerical full density matrix master equation simulations, taking into account all 8 relevant electronic and the lowest 17 motional states. An instance of this is presented in Fig.~\ref{fig:Time-dependent-rate}, where $R(\bar{n}_0, t)$ is shown as a function of the elapsed cooling time $t$ and initial mean population $\bar{n}_0$. $R(\bar{n}_0, t)$ falls as $\bar{n}_0$ increases, but it recovers as a function of time as lower thermal states become involved in the dynamics. This effect is connected with the observation of an increase of the $^2$P$_{1/2}$-state population when leaving the Lamb-Dicke regime. A possible explanation for this observation is that the strong laser-induced spin-motional coupling prevents establishing an electronic dark state that has no admixture of the excited $^2$P$_{1/2}$-state. Increasing $^2$P$_{1/2}$ population results in an enhanced scattering on carrier and sidebands, limiting the cooling rate and achievable $n_\mathrm{ss}$. The effect is more pronounced the higher Fock states are involved, but even at an $\bar{n}<1$ a \SI{20}{\%} reduction in the cooling rate can be observed (see Fig.~\ref{fig:Time-dependent-rate}). This behavior is characteristic of strong coupling dissipative theory \cite{Cerrillo2016}. A full understanding of this behavior requires the development of an analytical model for cooling in the regime beyond the LD approximation, which we leave to future theoretical investigations.

\subsection{Simulated data}
\label{subsec:simulated_data}
The simulated data is obtained by full density matrix theory as described in Appendix~\ref{master_equation}. For each data point the experimentally obtained $\Omega_\mathrm{\pi}$ and $\Delta$ are used in the simulation. In D-EIT cooling the Rabi frequencies of the pump beams are chosen such that the corresponding bright states coincide with the desired mode frequencies. In EIT cooling the simulated Rabi frequency of the \SI{866}{nm} beam is set to have no effect on the cooling other than clearing out the $^2$D$_{3/2}$ fast enough not to limit the cooling rate. The given simulated $\bar{n}$ are derived via sideband thermometry as in the experimental case. From these the cooling rates are derived to ensure comparability with the experimental data. The sideband thermometry leads to smaller $\bar{n}$ than would be obtained by averaging over all Fock states. However, that difference is not significant within the error bars of the measured $\bar{n}$. As discussed in this section, the cooling rates depend on the initial temperature of the cooled mode and change with elapsed cooling time, which is why it is not possible to define a single cooling rate for a given EIT cooling scheme. In order to be able to compare a simulated cooling rate to the experimental data, the simulated rate is given at $\bar{n}(t)=1$. 

%% Exp. results %%%%%%
\section{Results}\label{results}
In this section, the experimental results are presented and complemented by full density matrix simulations as described in Sec.~\ref{subsec:simulated_data} and Appendix~\ref{master_equation}. The scope of this work is the demonstration of D-EIT cooling of all three modes of a single $^{40}$Ca$^+$ ion in one cooling pulse and the comparison of its performance with the standard single EIT cooling technique. We first describe the implementation of single EIT cooling of the axial mode or both radial modes and provide resulting cooling rates and achievable $n_\mathrm{ss}$. Special emphasis is put on detrimental effects of single EIT cooling on the spectator modes for which the cooling pulse has not been optimized. Secondly, we demonstrate cooling of all three modes of a single $^{40}$Ca$^+$ ion by applying two consecutive single EIT cooling pulses, one to cool the axial mode and one to cool both close-by radial modes. Besides the cooling performance, the effect of the second cooling pulse on the initially cooled mode is investigated. This 3D single EIT cooling illustrates the possibilities and limitations of multi-mode cooling via standard EIT cooling. Finally, the results of D-EIT cooling are presented and compared to the single EIT cooling schemes and to the simulations.

In the following, $\bar{n}_\mathrm{i}$ represents the non steady-state mean motional quantum number and $n_\mathrm{ss,i}$ the steady-state mean motional quantum number, while $R_\mathrm{i}$ denotes the cooling rate. The index $i$ refers to either the axial (a) or the radial modes (r). 
\subsection{Single EIT cooling}
\begin{figure}
	\includegraphics[width=.99\linewidth]{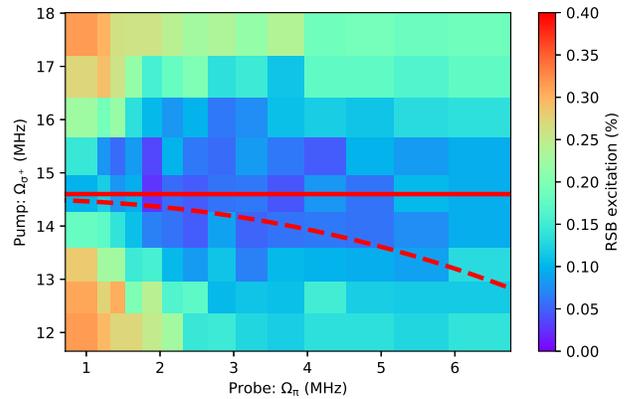}
	\caption{\label{Rabivspower}(color online) False-color map of the residual axial RSB excitation, depending on the Rabi frequencies of the pump and probe cooling beams after axial single EIT cooling. Areas of low RSB excitation correspond to pump/probe Rabi frequency combinations leading to low $\bar{n}_\mathrm{a}$. The horizontal line indicates the pump Rabi frequency along which the measurements shown in Fig.~\ref{Rabiscan} were taken. A constant combined Stark shift of the bright state corresponding to the axial trap frequency due to the pump and probe lasers is given the by red dashed line.}
\end{figure}
\begin{figure}
	\includegraphics[width=.99\linewidth]{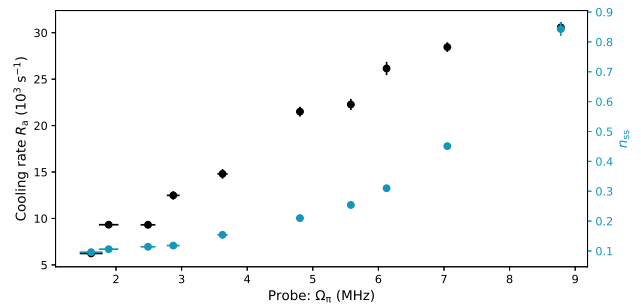}
	\caption{\label{Rabiscan}(color online) Dependency of $R_\mathrm{a}$ and $n_\mathrm{ss,a}$ on the probe Rabi frequency after axial single EIT cooling at $\Delta=3.4\:\Gamma$, while the pump Rabi frequency is kept fixed at \SI{14.6}{\mega\hertz}.}
\end{figure}

Fig.~\ref{Rabiscan} illustrates the dependence of $R_\mathrm{a}$ and $n_\mathrm{ss,a}$ on the probe Rabi frequency after axial single EIT cooling at $\Delta=3.4\Gamma$ and a fixed pump Rabi frequency. Low probe Rabi frequencies result in low $n_\mathrm{ss,a}$ and small $R_\mathrm{a}$. With increasing Rabi frequency the rate $R_\mathrm{a}$, but also $n_\mathrm{ss,a}$ increases due to off-resonant scattering via the unused excited state $|^2P_\mathrm{1/2},m_\mathrm{F}=-1/2\rangle$, as predicted by Eq.~\ref{eq:REIT}. Due to this trade-off, there is no single probe beam Rabi frequency $\Omega_{\pi}$ where both parameters are optimized. 

In order to account for the trade-off between $R$ and $n_\mathrm{ss}$ and to retrieve consistent results for single EIT cooling of a particular mode, the Rabi frequencies of the pump and probe beam are experimentally determined in a two-step process. In the first step, to find the Rabi frequencies of the cooling lasers that minimize the steady-state mode occupation $n_\mathrm{ss}$, we use a \SI{500}{\micro\second} cooling pulse long enough to reach steady-state for all reasonable cooling parameters. To maximize the cooling rate $R$ in the second step, we minimize $\bar{n}(t)$ after a \SI{100}{\micro\second} pulse, which is too short to reach steady-state. As a consequence, the combination of Rabi frequencies for pump and probe leading to the smallest RSB excitation is shifted towards higher probe Rabi frequencies, maximizing the cooling rate. The combined optimum of the two optimization steps was obtained from the overlapped Rabi frequency maps of the two steps, resulting in a compromise between a fast cooling rate and a low corresponding $n_\mathrm{ss}$.

Fig.~\ref{Rabivspower} shows an example of a map of the optimization of pump and probe Rabi frequencies for single axial EIT cooling with $\Delta=3.4\Gamma$. The colors represent the residual excitation on the axial RSB depending on the pump and probe Rabi frequencies after a fixed EIT cooling time of \SI{300}{\micro\second}, where blue indicates areas of low $n_\mathrm{ss}$. Fig.~\ref{Rabiscan} has been obtained by scanning the probe Rabi frequency along the red solid line indicated in Fig.~\ref{Rabivspower} for a fixed pump power of \SI{14.6}{\mega\hertz}.

\begin{figure}
	\includegraphics[width=.99\linewidth]{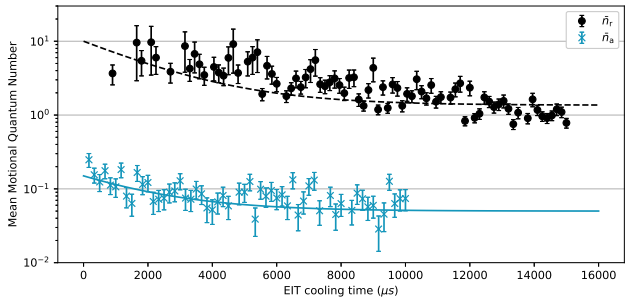}
	\caption{\label{slow_cooling}(color online) $\bar{n}(t)$ of the radial (black circles) and the axial (blue crosses) modes during single EIT cooling of the axial mode at $\Delta=3.4\:\Gamma$, $\Omega_\mathrm{\pi}=3.8\:$MHz, $\Omega_\mathrm{\sigma}=11.4\:$MHz, of the radial mode (black circles) and axial mode (blue crosses). The dashed and dotted lines represent exponential decay functions using the simulation results. Both curves employ the same cooling rate of \SI{439}{\per\second}, illustrating that the time to reach $n_\mathrm{ss,ax}$ is limited by the radial cooling rate. Radial mode: $n_\mathrm{ss}=1.36$, initial temperature $\bar{n}=10$. Axial mode: $n_\mathrm{ss}=0.05$, initial temperature of $\bar{n}=0.15$.}
\end{figure}
Each cooling pulse affects all three modes, not only the one for which it had been optimized. For each cooling pulse setting a cooling rate $R$ and a steady-state occupation $n_\mathrm{ss}$ can be assigned to every mode according to the resulting scattering spectrum as shown in the inset of Fig.~\ref{EITscat}. Generally, the bright resonance of single EIT cooling allows optimized cooling around one frequency, while modes further away experience smaller $R$ and higher $n_\mathrm{ss}$. The mode frequency range within which single EIT cooling to low $n_\mathrm{ss}$ is possible, is defined by the width of the bright state $\gamma_\mathrm{bright}\simeq\nu_\mathrm{r,a}\gamma/4\sqrt{\Delta^2+\Omega^2}$ \cite{Morigi_Cooling_2003}. E.g. for $\Delta=100\:$MHz, cooling of modes within a frequency range of $\approx 1.3\:$MHz to $n_\mathrm{ss}<0.1$ has been demonstrated in \cite{Lechner2016_EIT}. 

In the case of radial EIT cooling, the pump Rabi frequency required to shift the bright state into resonance with the radial mode frequency makes it broad enough to also provide fast cooling on the axial mode, although being located at the slope of the bright state. For our typical experimental parameters, the ratio of the axial and radial scattering rates multiplied by the ratio of Lamb-Dicke factors $\eta^2_\mathrm{a}/\eta^2_\mathrm{r}\approx2.8$ is approximately one, thus yielding similar cooling rates for the axial and radial modes. Conversely, for axial EIT cooling the bright resonance is narrower than for radial EIT cooling, since the pump Rabi frequency necessary to shift the bright resonance onto the axial mode is smaller. Here, the axial scattering rate is typically on the order of a factor of five larger than the radial scattering rate. Including the ratio of the Lamb-Dicke factors, the resulting radial cooling rate $R_\mathrm{r}$ is more than one order of magnitude slower than the axial cooling rate $R_\mathrm{a}$.

Lamb-Dicke and master equation predictions do not take into account the perturbation that the presence of an additional,
vibrationally hot degree of freedom (spectator modes \cite{wineland_experimental_1998, lin_sympathetic_2013}) generates in the cooling process. In general, one may regard any spectator modes as a source of dephasing that harms the cooling process. Therefore, the efficiency of cooling depends on the motional excitation of the spectator modes.

A nearby hot mode results in decoherence, indistinguishable from laser noise, via higher order scattering processes, which leads to an imperfect dark state and allows for weak carrier scattering \cite{wineland_experimental_1998}. As a consequence, the cooling rate of the cooled mode decreases while its $n_\mathrm{ss}$ increases compared to the case of a perfect dark state or cold spectator mode \cite{lin_sympathetic_2013}. In our setup, this cross-talk becomes relevant in axial EIT cooling, where $R_\mathrm{r}$ is much slower than $R_\mathrm{a}$. This behavior is illustrated in Fig.~\ref{slow_cooling}, which shows $\bar{n}_\mathrm{a}(t)$ and $\bar{n}_\mathrm{r}(t)$ during axial EIT cooling at $\Delta=3.4\:\Gamma$. After $\approx\:$\SI{500}{\micro\second} the axial mode is cooled to $\bar{n}_\mathrm{a} \approx n_\mathrm{ss,a}+0.1\approx0.15$, while the radial mode is still close to its initial temperature. Hence, the radial mode is hot enough to undermine the dark state and thus limits the $\bar{n}_\mathrm{a}$. For longer cooling times, $\bar{n}_\mathrm{a}$ decreases further at the rate $R_\mathrm{r}$, since the dark state becomes purer with decreasing $\bar{n}_\mathrm{r}$. This is supported by the dashed and dotted exponential decay functions for the radial and axial $\bar{n}(t)$, plotted together with the experimental data in Fig.~\ref{slow_cooling}, both using the same simulated $R_\mathrm{r}=439\:/$s, and the individual initial $\bar{n}$, and  $n_\mathrm{ss}$ during axial EIT cooling.

\begin{figure*}
	\includegraphics[width=1.\linewidth]{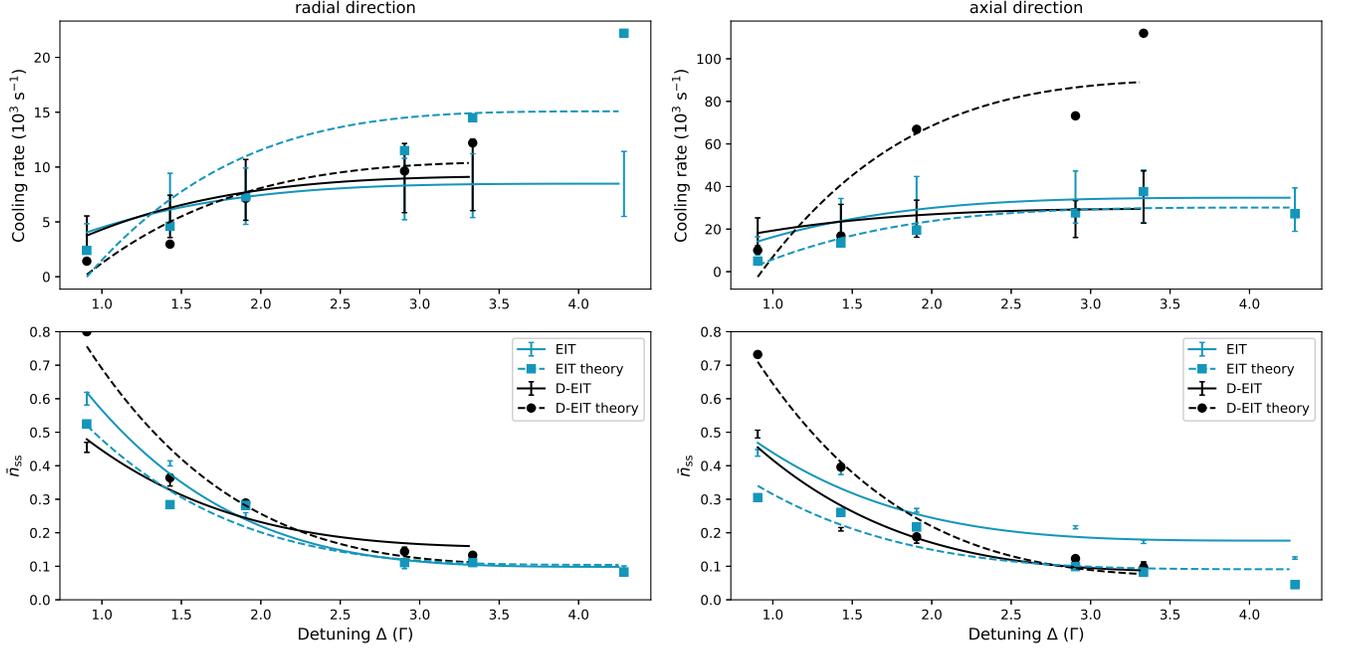}		
	\caption{(color online) Cooling rates and equilibrium $\bar{n}$ for EIT and D-EIT as a function of the blue detuning $\Delta$. (a) Radial modes. (b) Axial mode. The top panels display the cooling rates and the bottom panels the $n_\mathrm{ss}$. The solid and dashed lines are quartic polynomials fitted to the data points to guide the eye. The error bars incorporate systematic and statistical uncertainties. \label{expresults}}
\end{figure*}

The results of EIT cooling optimized individually for the axial and radial mode are shown in Fig.~\ref{expresults} together with simulated values for $R$ and $n_\mathrm{ss}$. Typical results of single EIT cooling are summarized in Tab.~\ref{EITcomparison}. It is impractical to wait for the axial mode to reach its final $n_\mathrm{ss}$. Therefore, in the case of axial EIT cooling, $\bar{n}_\mathrm{a}$ is measured after a time where it would have reached its equilibrium value if there was no cross-talk from hot radial modes. As a consequence, there is an offset of 0.1 between the simulated $n_\mathrm{ss,a}$ and the measured $\bar{n}_\mathrm{a}$. In accordance with the ratio of the Lamb-Dicke factors between the radial and axial mode, $R_\mathrm{a}\approx 2.8\times R_\mathrm{r}$, as can be seen from the top panels in Fig.~\ref{expresults}. For radial EIT cooling $n_\mathrm{ss,r}$ is well described by the full master equation theory. The simulated rates agree with the measured ones for $\Delta<3\Gamma$. For higher blue detunings the simulated radial rates are larger than the measured ones. This may be explained by cross talk from the axial mode, analogous to the cross-talk of the radial modes in the case of axial EIT cooling described above and shown in Fig.~\ref{slow_cooling}.
Although this effect cannot be captured by single-oscillator simulations, the Lamb-Dicke prediction for $R_\mathrm{a}$ within the radial single EIT cooling pulse is on the order of \SI{8e3}{\per\second}, which is similar to the measured $R_\mathrm{r}$. This serves as an indication that cross-talk from the axial mode is indeed limiting the $R_\mathrm{r}$, since departure from the Lamb-Dicke limit is ruled out by the coincidence between master equation simulations and Lamb-Dicke predictions.
\subsection{3D single EIT cooling}
Simultaneous cooling of different mode frequencies within a single EIT cooling pulse is limited by the finite width of the bright state  (see Fig.~\ref{EITscat}, \cite{Lechner2016_EIT, lin_sympathetic_2013}). In order to provide simultaneous cooling of mode frequencies that lie outside the single EIT cooling bandwidth, several EIT pulses have to be applied sequentially, each of them optimized to cool around one particular frequency, in our case the axial or radial mode(s). In the following, the first cooling pulse is termed \emph{precooling pulse}. It is optimized to cool one of the modes (precooled mode) with high cooling rate and aims to reach a low $n_\mathrm{ss}$. During that pulse the off-resonant mode(s) are cooled as well, but at a slower rate and to a higher $\bar{n}$, as demonstrated in Fig.~\ref{slow_cooling}. To keep the overall cooling time short, the precooling pulse is terminated once the precooled mode reached its $n_\mathrm{ss}$, although the other mode(s) might not have reached their steady state, yet. After that, a second pulse optimized for cooling the other mode(s) is applied. 
\begin{figure}
	\includegraphics[width=.99\linewidth]{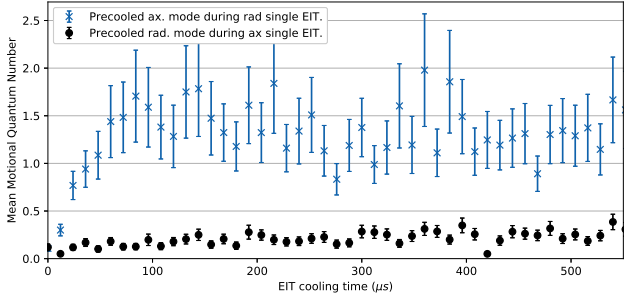}
	\caption{\label{precheating}(color online) Temperature of the precooled mode during the second cooling pulse of 3D single EIT cooling. For long enough cooling times of the second cooling pulse, both precooled modes will approach $n_\mathrm{ss}\approx1.3$.}
\end{figure}

A cooling pulse tailored for a certain mode may detrimentally affect other modes. As one can conclude from the analysis of typical cooling spectra $S\left(\omega\right)$, away from a bright resonance the ratio $\frac{S\left(\nu\right)}{S\left(-\nu\right)}$ is not favorable and steady-state occupation numbers above $1$ are expected. Therefore, it is generally the case that a concatenation of pulses may result in less effective cooling.

The $n_\mathrm{ss}$ of the precooled mode associated with the scattering spectrum of the second cooling pulse is higher than the $n_\mathrm{ss}$ at the end of the precooling pulse and it will be heated up, as depicted in Fig.~\ref{precheating}. It shows the increase of $\bar{n}(t)$ of the axial and radial precooled mode caused during the second cooling pulse. This effective heating is smaller for the precooled radial mode than for the precooled axial mode. Thus, the ordering of the two cooling pulses leads to different $\bar{n}$ for both modes after the second cooling pulse, see also Table~\ref{EITcomparison}.

Typical results of 3D EIT cooling are summarized in Tab.~\ref{EITcomparison}. The listed $\bar{n}$ of the precooled mode depends on the cooling time of the second mode. Since the precooling pulse cools the second mode below Doppler temperature, the time for the second mode to reach $n_\mathrm{ss}$ is reduced. Axial EIT with radial precooling is the most beneficial implementation of 3D single EIT cooling in our setup. All three modes of a single $^{40}$Ca$^+$ ion are cooled close to the ground state within \SI{640}{\micro\second} combined cooling time.
\subsection{D-EIT cooling}
Fig.~\ref{expresults} displays and compares the results of single EIT and D-EIT cooling. For each method the $n_\mathrm{ss}$ and cooling rate for the radial and axial modes is shown as a function of the blue detuning $\Delta$ of the virtual $P_\mathrm{+}^v$ from the magnetic field-free $^2S_{1/2}$ to $^2P_{1/2}$ transition. The maximum detuning is technically limited by the tuning range of the \SI{866}{nm} AOM in case of D-EIT and by the \SI{397}{nm} AOM in case of single EIT cooling. The optimal cooling parameters were found experimentally by adjusting the three involved Rabi frequencies to obtain ac Stark shifts of the bright states matching the two mode frequencies, while maximizing the cooling rate and minimizing the steady-state $n_\mathrm{ss}$.

An increasing detuning $\Delta$ yields a higher cooling rate -- due to higher Rabi frequencies of the cooling beams -- and a lower steady-state occupation $n_\mathrm{ss}$. The latter is a consequence of an improved ratio between cooling and heating scattering events, since the bright state scattering resonances narrow  while the residual blue sideband excitations are suppressed. D-EIT achieves cooling rates and steady-state energies similar to those obtained from single EIT cooling, but does so for all modes of a $^{40}$Ca$^+$ ion's motion simultaneously, while single EIT cooling had to be optimized for GSC of each mode separately.

The simulated $n_\mathrm{ss}$ as well as the radial cooling rates agree well with the measured values. In the axial direction the simulated cooling rates are higher than the measured ones. This difference increases with $\Delta$ and may be explained by the limited maximum number of Fock states that can be numerically simulated. The truncation in the simulation of higher Fock states with non-negligible occupation leads to an overestimation of $R_\mathrm{a}$ \cite{chen_sympathetic_2017}. We have verified, that $R_\mathrm{a}$ in D-EIT decreases with increasing number of simulated Fock states. This effect is more pronounced in the axial direction when the lower trap frequency results in population of larger Fock states compared to the radial direction for the same initial temperature. Furthermore, we suspect that high-frequency relative phase noise between the \SI{866}{\nano\meter} and \SI{397}{\nano\meter} lasers results in a deterioration of the dark state, further restricting the experimentally observed cooling rate at high Rabi frequencies.

\begin{table}
	\caption{\label{EITcomparison} Comparison of experimental values for single EIT, single EIT with precooling and D-EIT cooling from initial Doppler cooling temperatures, each optimized for minimum $\bar{n}$ at $\Delta=3.4\:\Gamma$. In case of axial single EIT cooling the non-steady-state $\bar{n}$ is given (cf. Fig.~\ref{slow_cooling}).}
	\begin{ruledtabular}
		\begin{tabular}{lccc}
			EIT Method 	& $n_\mathrm{ss,a}$ & $n_\mathrm{ss,r}$ & $t_\mathrm{ss}$ ($\mu$s)\\
			\hline 	
			Single EIT ax. & 0.10 & $\approx$7 & 520\\
			Single EIT ax. w/ rad. prec. &	0.04 & 0.20 & 640\footnotemark[1]\\
			Single EIT rad. & $\approx$1.3 & 0.10 & 490\\
			Single EIT rad. w/ ax. prec. & $\approx$1 & 0.10 & 1120\\
			D-EIT & 0.11 & 0.14 & 670\\	
		\end{tabular}
	\end{ruledtabular}
	\footnotetext[1]{Smaller than the sum of $t_\mathrm{ss}$ since the axial mode is pre-cooled by the radial pre-cooling pulse.}
\end{table}
Table \ref{EITcomparison} compares the final $n_\mathrm{ss}$ and the time $t_\mathrm{ss}$ needed to reach the final temperature between D-EIT cooling and the results of single EIT cooling in axial and radial direction with and without precooling. These numbers were obtained from measurements at $\Delta=3.4\:\Gamma$, which is the maximum blue detuning usable for both single and D-EIT cooling in our apparatus. With the parameters employed here, the cooling performance of D-EIT is comparable to that of precooled axial single EIT cooling. 

In our experiment, standard EIT cooling is 15 times faster than SBC \cite{roos_quantum_1999} and still roughly seven times faster than the fastest ever reported SBC of a single $^{40}$Ca$^+$ ion of which we are aware \cite{huber_quantum_2010}. Since the observed cooling rates of D-EIT cooling are similar to that of single EIT cooling, D-EIT cooling is also faster compared to SBC.
\subsection{Comparison between Lamb-Dicke analytical theory, master equation simulation and experiment}
\begin{figure}
	\includegraphics[width=1.\linewidth]{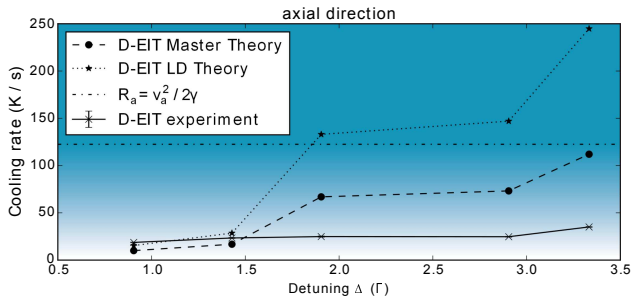}		
	\caption{(color online) Axial cooling rates for D-EIT as a function of the blue detuning $\Delta$. Compared are the measured rates with the simulated values obtained by the full master equation approach with limited Fock states and the Lamb-Dicke approximation. The dash-dotted line indicates the limit for the validity of the Lamb-Dicke approximation, see Eq.~\ref{eq:LDlimit}, the background gradient represents the gradual failure of the Lamb-Dicke approximation.  \label{rate_master_vs_LD}}
\end{figure}

Fig.~\ref{rate_master_vs_LD} illustrates the breakdown of the simple analytical model in the Lamb-Dicke approximation. As detuning increases, the cooling rates predicted by Eq.~\ref{eq:REIT} grow an order of magnitude faster than those observed in the experiment. Some of the discrepancy is due to the correlations between electronic and vibrational degrees of freedom captured by the master equation simulations. The simulations, accordingly, come closer to the experimentally observed behavior. It is not yet clear whether the remaining discrepancy is due to numerical limitations or to other features not captured by our model such as relative phase noise between the cooling lasers.

\section{Conclusion and outlook}\label{conclusion}
In conclusion, we developed double-bright EIT (D-EIT) cooling as a novel approach to EIT cooling. D-EIT allows simultaneous ground-state cooling of very different mode frequencies, which is experimentally demonstrated by GSC of all three motional degrees of freedom of a single $^{40}$Ca$^+$ ion within one cooling pulse. We compared the steady-state mean motional quantum numbers and cooling rates of D-EIT to those of single EIT cooling and 3D EIT cooling both experimentally and theoretically.

We give a thermodynamic interpretation of laser cooling by using an analogy between linear response theory and the Lamb-Dicke regime. Within the Lamb-Dicke regime, a description of the system in form of an analytical theory for both single EIT and D-EIT cooling is given. A theory for time-dependent rates has been developed to describe the system beyond the Lamb-Dicke regime and it has been compared to the measured data. We demonstrate how the frequency range of single EIT cooling is limited by the finite width of the cooling resonance  and can heat other motional degrees of freedom that are outside the cooling bandwidth. Furthermore, the detrimental effect of spectator modes on the $n_\mathrm{ss}$ of the cooled mode is observed. In the $^{40}$Ca$^+$ level structure, cooling rates for single EIT cooling are limited by off-resonant scattering via the unused magnetic sublevel of the $P_\mathrm{1/2}$ state. The chosen level scheme of D-EIT cooling protects the dark state from such decoherence processes as well as limitations of cooling rates and $\bar{n}$ due to motional spectator modes. Thus, for systems with smaller Lamb-Dicke factors, it can reach higher cooling rates and lower $n_\mathrm{ss}$ and outperform single EIT cooling. 

The D-EIT technique is attractive for e.g. multi-ion, multi-species experiments in which quantum control of several spectrally separated motional modes is required. Examples include ion strings to study many-body physics \cite{schneider_experimental_2012, Kiethe_Probing_2017}, quantum simulations \cite{porras_effective_2004, blatt_quantum_2012}, and quantum information processing \cite{kienzler_observation_2016}. The technique will be particularly useful for quantum logic spectroscopy \cite{wan_precision_2014, wolf_non-destructive_2016, rosenband_frrequency_2008, Schmidt2005, hempel_entanglement-enhanced_2013, chou_preparation_2017}, and quantum logic clocks \cite{rosenband_frrequency_2008, Ludlow_ClocksReview} for which two ions of a different species need to be cooled to suppress systematic frequency shifts \cite{chen_sympathetic_2017}. In our experiment, a single $^{40}$Ca$^+$ ion is used to sympathetically cool a single $^{27}$Al$^+$ clock ion and serves as the logic ion in quantum logic spectroscopy \cite{Schmidt2005}. Due to their mass ratio and the high radial symmetry of our trap, it is possible to choose a set of experimental parameters such that all six motional modes of the two ion crystal bunch at two frequencies. By rotating the trap by $45^\circ$ in the horizontal (xz) plane with respect to the k-vectors of the cooling beams, D-EIT will enable ground state cooling of all six modes of the two ion-crystal in one single cooling pulse. This helps to reduce the dead-time, and thus reduces the instability of the clock. In ion traps with large heating rates, near-resonant D-EIT cooling would establish a steady-state $n_\mathrm{ss}$ below the Doppler cooling limit with low incident power levels to minimize ac-Stark shifts on the clock transition \cite{rosenband_frrequency_2008}. Furthermore, the D-EIT cooling scheme can provide even triple EIT cooling if the magnetic field strength is chosen such that the $|^2D_{3/2}, m_\mathrm{F}=\mp1/2\rangle$ sub-level is red detuned from the virtual excited state $P_+^v$ by the frequency of the radial mode and if the \SI{397}{\nano\meter} is $\sigma^\pm$ polarized. That way, the D-EIT scheme additionally provides a second dark state at the position of the radial motional blue sideband, suppressing all second order radial heating processes. In principle, D-EIT could be extended to cool more modes by the creation of additional bright states through further lambda-like resonances connecting sub-levels of the $^2S_{1/2}$, $^2D_{3/2}$, or $^2D_{5/2}$ states via the $^2P_{1/2}$ or $^2P_{3/2}$ manifolds. This scalability is limited by the number of such possible combinations.
\begin{acknowledgments}
	The authors would like to thank Nicolas Spethmann for helpful comments on the manuscript. We acknowledge support from the DFG through CRC 1128 (geo-Q), project A03 and CRC 1227 (DQ-mat), project B03, and the state of Lower Saxony, Hannover, Germany. 
\end{acknowledgments}

\appendix
\section{Master Equation}
\label{master_equation}

	The simulations have been performed with a master equation including
	the 8 electronic levels ($\left|S_{+}\right\rangle ,\left|S_{-}\right\rangle ,\left|P_{+}\right\rangle $, $\left|P_{-}\right\rangle ,\left|D_{+3}\right\rangle ,\left|D_{+}\right\rangle ,\left|D_{-}\right\rangle ,\left|D_{-3}\right\rangle $)
	and one vibrational degree of freedom represented by the annihilation
	operator $b$. We assume the presence of 397 nm laser in $\pi$ and
	$\sigma^{+}$ polarizations and 866 nm in $\sigma^{+}$ and $\sigma^{-}$
	polarizations. The Hamiltonian in $\hbar$ units is 
\begin{widetext}
	\begin{alignat*}{1}
	H & =\nu b^{\dagger}b+\Delta_{\pi}\left(\left|S_{+}\right\rangle \left\langle S_{+}\right|-\left|P_{-}\right\rangle \left\langle P_{-}\right|-\left|D_{+}\right\rangle \left\langle D_{+}\right|-\left|D_{-3}\right\rangle \left\langle D_{-3}\right|\right)\\
	& +\Delta_{\sigma}\left(\left|S_{-}\right\rangle \left\langle S_{-}\right|+\left|P_{-}\right\rangle \left\langle P_{-}\right|+\left|D_{+}\right\rangle \left\langle D_{+}\right|+\left|D_{-3}\right\rangle \left\langle D_{-3}\right|\right)\\
	& +\Delta_{D}\left(\left|D_{-3}\right\rangle \left\langle D_{-3}\right|+\left|D_{-}\right\rangle \left\langle D_{-}\right|+\left|D_{+}\right\rangle \left\langle D_{+}\right|+\left|D_{+3}\right\rangle \left\langle D_{+3}\right|\right)\\
	& +\mu_{B}B\left[\left|S_{+}\right\rangle \left\langle S_{+}\right|-\left|S_{-}\right\rangle \left\langle S_{-}\right|+\frac{1}{3}\left(\left|P_{+}\right\rangle \left\langle P_{+}\right|-\left|P_{-}\right\rangle \left\langle P_{-}\right|\right)\right.\\
	& \qquad\qquad\left.+\frac{2}{5}\left(3\left|D_{+3}\right\rangle \left\langle D_{+3}\right|+\left|D_{+}\right\rangle \left\langle D_{+}\right|-\left|D_{-}\right\rangle \left\langle D_{-}\right|-3\left|D_{-3}\right\rangle \left\langle D_{-3}\right|\right)\right]\\
	& +\frac{1}{2}\left(\Omega_{\pi}e^{i\eta_{\pi}\left(b+b^{\dagger}\right)}\left|S_{-}\right\rangle \left\langle P_{-}\right|+\Omega_{\pi}e^{i\eta_{\pi}\left(b+b^{\dagger}\right)}\left|S_{+}\right\rangle \left\langle P_{+}\right|+\Omega_{\sigma}e^{i\eta_{\sigma}\left(b+b^{\dagger}\right)}\left|S_{-}\right\rangle \left\langle P_{+}\right|+H.c.\right)\\
	& +\frac{\Omega_{D}}{2}\left[e^{i\eta_{D}\left(b+b^{\dagger}\right)}\left(\sqrt{3}\left|D_{-3}\right\rangle \left\langle P_{-}\right|+\left|D_{-}\right\rangle \left\langle P_{+}\right|+\left|D_{+}\right\rangle \left\langle P_{-}\right|+\sqrt{3}\left|D_{+3}\right\rangle \left\langle P_{+}\right|\right)+H.c.\right],
	\end{alignat*}

	where $\Delta_{\pi}$, $\Delta_{\sigma}$ and $\Delta_{D}$ correspond
	to the detunings of the laser frequencies with respect to the unperturbed
	transitions of the 397 nm laser in $\pi$ and $\sigma^{+}$ polarizations
	and 866 nm laser respectively. Their associated Rabi frequencies and
	Lamb-Dicke parameters are represented by $\Omega_{i}$ and $\eta_{i}$
	and the appropriate subindex $i$. The trap frequency is $\nu$, the
	magnetic field is $B$ and $\mu_{B}$ represents the Bohr magneton.
	The decay of the excited states is incorporated in the master equation
	of the density matrix $\rho$

	\begin{alignat}{1}
	\frac{d}{dt}\rho\left(t\right)=\mathcal{L}\rho\left(t\right) & =-i\left[H,\rho\right]+\sum_{j\in\left\{ S_{\pm},D_{\pm},D_{\pm3}\right\} }\sum_{i\in\pm}\Gamma_{j}\left(2\left|j\right\rangle \left\langle P_{i}\right|\rho\left|P_{i}\right\rangle \left\langle j\right|-\left|P_{i}\right\rangle \left\langle P_{i}\right|\rho-\rho\left|P_{i}\right\rangle \left\langle P_{i}\right|\right),\label{eq:ME-1}
	\end{alignat}
	where $\Gamma_{j}$ represents the decay of each dipole.

\section{Calculation of D-EIT Scattering spectrum}
\label{Supplementary_scattering}
The Hamiltonian for D-EIT for the electronic degrees of freedom and
vanishing Lamb-Dicke parameters is

\begin{alignat*}{1}
H'_{i} & =\Delta+\frac{\Delta_{s}}{2}-\Delta\left(\left|P_{+}\right\rangle \left\langle P_{+}\right|+\left|P_{-}\right\rangle \left\langle P_{-}\right|\right)\\
& +\Delta_{\mathrm{s}}\left[-\frac{1}{3}\left|P_{+}\right\rangle \left\langle P_{+}\right|-\frac{2}{3}\left|P_{-}\right\rangle \left\langle P_{-}\right|-\frac{1}{5}\left(3\left|D_{+3}\right\rangle \left\langle D_{+3}\right|+7\left|D_{-}\right\rangle \left\langle D_{-}\right|+4\left|D_{-3}\right\rangle \left\langle D_{-3}\right|\right)\right]\\
& +\frac{1}{2}\left[\left(\Omega_{\pi}\left|S_{-}\right\rangle +\sqrt{3}\Omega_{D}\left|D_{-3}\right\rangle +\Omega_{D}\left|D_{+}\right\rangle \right)\left\langle P_{-}\right|+\left(\Omega_{\sigma}\left|S_{-}\right\rangle +\Omega_{\pi}\left|S_{+}\right\rangle +\Omega_{D}\left|D_{-}\right\rangle +\sqrt{3}\Omega_{D}\left|D_{+3}\right\rangle \right)\left\langle P_{+}\right|+H.c.\right].
\end{alignat*}

The associated effective Hamiltonian is
\[
H_\mathrm{eff}=H'_{i}-i\gamma\left(\left|P_{+}\right\rangle \left\langle P_{+}\right|+\left|P_{-}\right\rangle \left\langle P_{-}\right|\right),
\]
with $\gamma$ the sum over all decay rates from a given excited state,
and the Lamb-Dicke coupling operator has now the form 
\[
\hat{\sigma}_{\eta}\equiv\frac{1}{2}\left[i\left(\eta_{\pi}\Omega_{\pi}\left|S_{-}\right\rangle +\eta_{D}\Omega_{D}\left|D_{+}\right\rangle \right)\left\langle P_{-}\right|+i\left(\eta_{\sigma}\Omega_{\sigma}\left|S_{-}\right\rangle +\eta_{\pi}\Omega_{\pi}\left|S_{+}\right\rangle \right)\left\langle P_{+}\right|+H.c.\right],
\]
where only terms with overlap to the dark state have been kept. The relevant
product is then 
\begin{alignat*}{1}
\sigma_{\eta}\left|\sim\right\rangle  & =\frac{-i}{2}\frac{\Omega_{\pi}\Omega_{D}\Omega_{\sigma}}{\Omega_{-}}\left(\eta_{\pi}-\eta_{\sigma}\right)\left|P_{+}\right\rangle +\frac{i}{2}\frac{\Omega_{\pi}^{2}\Omega_{D}}{\Omega_{-}}\left(\eta_{\pi}-\eta_{D}\right)\left|P_{-}\right\rangle \\
& =\frac{-i}{2}\frac{\Omega_{\pi}\Omega_{D}}{\Omega_{-}}\left[\Omega_{\sigma}\left(\eta_{\pi}-\eta_{\sigma}\right)\left|P_{+}\right\rangle +\Omega_{\pi}\left(-\eta_{\pi}+\eta_{D}\right)\left|P_{-}\right\rangle \right]\\
& \equiv-i\left|E\right\rangle ,
\end{alignat*}

where we have introduced the definition of the normalized state 
\[
\left|E\right\rangle \equiv\frac{\Omega_{\pi}\Omega_{D}}{2\Omega_{-}}\left[\Omega_{\sigma}\left(\eta_{\pi}-\eta_{\sigma}\right)\left|P_{+}\right\rangle +\Omega_{\pi}\left(\eta_{D}-\eta_{\pi}\right)\left|P_{-}\right\rangle \right].
\]
Due to the form of $\left|E\right\rangle $, it is possible to represent
the solution in terms of the fraction associated to the effective
Hamiltonian
\[
S\left(\omega\right)=\left\langle E\right|\cfrac{1}{i\omega+iH_{EE}-H_{EG}\cfrac{1}{i\omega-iH_{GG}}H_{GE}}\left|E\right\rangle ,
\]
where
\begin{alignat*}{1}
H_{EE} & =\left(\begin{array}{cc}
-\Delta-\frac{1}{3}\Delta_{\mathrm{s}}-i\gamma & 0\\
0 & -\Delta-\frac{2}{3}\Delta_{\mathrm{s}}-i\gamma
\end{array}\right),\\
H_{EG} & =\frac{1}{2}\left(\begin{array}{cccccc}
\Omega_{\pi} & \Omega_{\sigma} & \sqrt{3}\Omega_{D} & 0 & \Omega_{D} & 0\\
0 & \Omega_{\pi} & 0 & \Omega_{D} & 0 & \sqrt{3}\Omega_{D}
\end{array}\right),\\
H_{GE} & =H_{EG}^{\dagger},\\
H_{GG} & =\frac{\Delta_{\mathrm{s}}}{5}\times \mathrm{diag}\left(\begin{array}{cccccc}
0 & 0 & 3 & 0 & 7 & 4\end{array}\right).
\end{alignat*}
Therefore
\[
S\left(\omega\right)=-i\left\langle E\right|\left(\begin{array}{cc}
a\left(\omega\right) & -\frac{\Omega_{\pi}\Omega_{\sigma}}{4\omega}\\
-\frac{\Omega_{\pi}\Omega_{\sigma}}{4\omega} & b\left(\omega\right)
\end{array}\right)^{-1}\left|E\right\rangle .
\]
\end{widetext}

\bibliography{Bibliography/Main}

\end{document}